\documentclass{article}
\usepackage[a4paper,left=0.8in,right=0.8in,top=1in,bottom=1in]{geometry}
\usepackage{natbib}
\usepackage{mathrsfs}
\usepackage{amsmath,amsfonts,amssymb,mathrsfs}
\usepackage[all]{xy} % for diagrams
\usepackage{mathtools}
\usepackage{framed}
\usepackage{lineno} %[mathlines,displaymath]
\usepackage{bm}%bold
\usepackage{graphicx}
\usepackage{booktabs}
\usepackage{threeparttable}
\usepackage{tabularx}
\usepackage{authblk}

\graphicspath{{./art/}}
\newtheorem{proposition}{Proposion}
\newtheorem{lemma}{Lemma}
\newtheorem{theorem}{Theorem}
\newtheorem{assumption}{Assumption}
\newtheorem{define}{Definition}

\newcommand{\mbE}{\mathbb{E}}

\newcommand{\mbI}{\mathbb{I}}

\newcommand{\mbN}{\mathbb{N}}

\newcommand{\mbP}{\mathbb{P}}

\newcommand{\mbR}{\mathbb{R}}
\newcommand{\mbS}{\mathbb{S}}

\newcommand{\mbZ}{\mathbb{Z}}

\newcommand{\mcE}{\mathcal{E}}
\newcommand{\mcF}{\mathcal{F}}

\newcommand{\mcH}{\mathcal{H}}

\newcommand{\mcN}{\mathcal{N}}

\newcommand{\mcT}{\mathcal{T}}

\newcommand{\bW}{\bm{W}}

\newcommand{\bx}{\bm{x}}

\def\bbeta{\boldsymbol{\beta}}
\def\bmu{\boldsymbol{\mu}}

\def\bDelta{\boldsymbol{\Delta}}

\def\blambda{\boldsymbol{\lambda}}

\def\tr{\mathrm{T}}

\newcommand{\Cov}{\mathrm{Cov}}

\newcommand{\argmin}{\mathop{\arg\min}}

\newcommand{\supp}{\operatorname{supp}}

\renewcommand{\d}{\text{d}}

\title{Temporal Point Process Graphical Models}

\author{Yalong Lyu\thanks{The first and second authors have equal contributions to this paper.}, Huiyuan Wang and Wei Lin \thanks{weilin@math.pku.edu.cn}} 
\affil{School of Mathematical Sciences, Peking University, China}

\begin{document}

\maketitle

\begin{abstract}
Many real-world objects can be modeled as a stream of events on the nodes of a graph. In this paper, we propose a class of graphical event models named temporal point process graphical models for representing the temporal dependencies among different components of a multivariate point process. In our model, the intensity of an event stream can depend on the historical events in a nonlinear way. We provide a procedure that allows us to estimate the parameters in the model with a convex loss function in the high-dimensional setting. For the approximation error introduced during the implementation, we also establish the error bound for our estimators. We demonstrate the performance of our method with extensive simulations and a spike train data set.
% Please include a maximum of seven keywords

keywords: {graphical models; temporal point processes; high-dimensionality}
\end{abstract}

\section{Introduction}
\label{SectionIntroduction}
Graphical modeling of point processes is drawing increasing attention, as data in the form of point processes on a graph are emerging in scientific and business applications. Examples include biological neural networks with neural spike train data (see e.g. \citeauthor{apppaninski2007statistical}, \citeyear{apppaninski2007statistical}; \citeauthor{apppillow2008spatio}, \citeyear{apppillow2008spatio}; \citeauthor{2018Rec}, \citeyear{2018Rec}); social media data recording timestamps of each user's actions (\citeauthor{perry2013point}, \citeyear{perry2013point}; \citeauthor{zhou2013learning}, \citeyear{zhou2013learning}; \citeauthor{fox2016modeling}, \citeyear{fox2016modeling}); high frequency financial data recording times of market orders (\citeauthor{bacry2013some},\ \citeyear{bacry2013some}; \citeauthor{ait2015modeling}, \citeyear{ait2015modeling}; \citeauthor{bacry2015hawkes}, \citeyear{bacry2015hawkes}). Understanding the connectivity structure of these point processes, i.e, the edges on the graphs, is a challenging problem of growing interest. Our work is motivated by a recent spike train data set from \citeauthor{2018Rec} (\citeyear{2018Rec}) . This data set collected spike train data from two important brain regions for olfactory of mice, namely the olfactory blub and the piriform cortex. The neural activity in the olfactory bulb is highly concentration-dependent, while that in the piriform cortex is concentration-invariant. The study of \citeauthor{2018Rec} (\citeyear{2018Rec}) suggests that there exists a complex excitation-inhabitation mechanism between the olfactory blub and the piriform cortex that allows the nervous system to produce a concentration-invariant representation of odors.

In this article, we propose a new temporal point process graphical model to address this problem. We model the spike train data of the neurons as a multivariate temporal point process, where each component of the point process represents the activity of a neuron. We allow future events be excited or inhabited by the past events in a non-additive manner. We organize the transfer coefficients into a matrix and impose a sparse structure, which is welcomed for interpretation in neuroscience and other applications (\citeauthor{zhao2006model}, \citeyear{zhao2006model};
\citeauthor{zou2006adaptive}, \citeyear{zou2006adaptive}). Our loss function is convex, so many highly scalable algorithms, such as alternating direction method of multipliers and coordinate descent method, can be applied to parameter estimation. We establish a non-asymptotic error bound of the estimator and allow the number of nodes in the graph to diverge. We also analyze the approximation error introduced during the implementation phase.

There have been a great number of works focusing on graphical models, including the Gaussian graphical model (\citeauthor{Gaussianyuan2007model}, \citeyear{Gaussianyuan2007model};
\citeauthor{Gaussianliu2013gaussian}, \citeyear{Gaussianliu2013gaussian}) and directed acyclic graph (\citeauthor{DAGkalisch2007estimating}, \citeyear{DAGkalisch2007estimating};
\citeauthor{DAGvandegeer2013}, \citeyear{DAGvandegeer2013};
\citeauthor{DAGzheng2018dags}, \citeyear{DAGzheng2018dags}). In this models, each node of the graph is corresponding to a random variable and the data are realizations of the random variables. The graphs of classical graphical models are used to represent the structure of conditional independence among the random variables. In contrast, we aim at the case where each node of the graph represents a point process and the graph can be viewed as a multivariate point process. In our setting, the graph encodes the \textit{local independence} relationship among the components of the multivariate point process (\citeauthor{didelez2008graphical}, \citeyear{didelez2008graphical}). The basic idea of local independence is that once we know about specific past events, the intensity of the future event is independent of other past events. In our model, local independence is fully characterised by the parameters (see Proposition \ref{Propositionlocalindependence}).

There have been a family of models targeting temporal point processes. A Hawkes process (\citeauthor{1971Hawkes}, \citeyear{1974Hawkes}) assumes that historical events can trigger future events and has been widely used in modelling neuronal spike train data and social network data. The model we proposed differs from existing research of the Hawkes process in many ways. Most prior art has focus on parameter estimation in the linear case, where the transfer function of the multivariate Hawkes process is the identity function. \citeauthor{2015Hansen}(\citeyear{2015Hansen}) considered estimating parameters in a linear Hawkes process of fixed number of nodes with least-square loss function and $\ell_1$ regularization. \citeauthor{bacry2020sparse}(\citeyear{bacry2020sparse}) organized the transferring coefficients as a matrix and impose a low-rank structure. A crucial weakness of linear Hawkes process is that it does not allow inhibitory relations, i.e, the past events can only trigger more future events, not silence future events. Our model extends to a nonlinear setting where inhibitory influences are allowed and the influence of past events from different nodes can accumulate in a non-additive manner.

\citeauthor{tang2021multivariate}(\citeyear{tang2021multivariate}) also considered parameter estimation of a nonlinear Hawkes process in their work. Our work differs from \citeauthor{tang2021multivariate}(\citeyear{tang2021multivariate}) in several aspects. First, we focus on the estimation of sparse graphical models while they focus on the estimation of a low-rank tensor of regression parameters. Second, we provide a stronger theoretical guarantee by proving the global minimum of our penalized loss function has a $\sqrt{\frac{s\log p}{T}}$ convergence rate while their theory only shows that there exists a local minimum of the penalized loss function in the neighborhood of the true parameter. Finally, our loss function is convex while theirs is not. A convex loss function with restricted strong convexity allows us to find global minimum efficiently.

The paper is organized as follows. We will finish this section with notation. We briefly introduce the temporal point process and our model in Section 2. In Section 3, we present the construction of our estimator and our main theorem. In Section 4, we analyze the approximation error introduced during the implementation phase. We show the results of simulations in Section 5 and an application on a spick train data set in Section 6. We conclude with a discussion in Section 7. Technical proofs are provided in the Appendix.

Some notations will be employed through out this paper. Let $[n], n \in \mbN $ denote the set $\{1,\ldots, n\}$. For a set $S \subset [p]$ and a vector $v \in \mbR^p$,
we define $v_S \in \mbR^p$ as $(v_S)_i = v_i \mbI(i\in S)$. Let $\|\cdot\|_p,\ p \in [1,\infty)$ denote the $\ell_p$-norm of a vector and $\|\cdot\|_0$
denote the number of non-zero entries of a vector. For a matrix $A \in \mbR^{m\times n}$, let $\|A\|_F = (\sum_{i=1}^m\sum_{j=1}^n a_{ij}^2)^{1/2}$ denote the Frobenius norm of a matrix. We use $S_q^p(r)$ for a $\|\cdot\|_q$ ball with radius $r$ in $\mbR^p$. We use $c_1,c_2,\ldots$ to denote positive numbers that do not depend on $p$ or $T$ but can still depend on some values that we consider as constants in this paper, and their value can change from line to line.
\section{The Temporal Point Process Graphical Model}
We will begin with a very brief review of temporal point processes. For a more systematic and comprehensive discussion of point processes, we refer to \citeauthor{Pierre1981Point} (\citeyear{Pierre1981Point}) and \citeauthor{daley2007introduction} (\citeyear{daley2007introduction}).

Let $N(t), t\geq 0$ be a simple point process. $N(t)$ is corresponding to a sequence of real-valued random variables $\{t_k\}_{k\in\mbZ}$, with $t_{k+1} > t_k$ and $t_1 \geq 0$ almost surely. $\{t_k\}_{k\in\mbZ}$ are called the event times of $N(t)$. For a set $A$ in the Borel $\sigma$-field of the real line, $N(A) = \sum_{k}\mbI_{\{t_k\in A\}}$. We write $N([t,t + \mathrm{d} t))$ as $\mathrm{d} N(t)$, where $\mathrm{d} t$ denotes an arbitrarily small increment of $t$. Let $\mcH_t$ denote the $\sigma$-field generated by $N(\tau), \tau \in [0,t]$. The intensity of $N(t)$ is defined as
$$\lambda(t) \mathrm{d} t = \mbP(\mathrm{d} N(t) = 1|\mcH_t).$$

Now suppose we have $p$ simple point processes $N_1,\ldots, N_p$ with $t_{j,1} < t_{j,2} <\cdots < t_{j,n} <\cdots$ denoting the event times of the process $N_j, j \in [p]$. Then we employ $\mcN = (N_1,\ldots,N_p)$ to denote a $p$-variate point process. Let $\mcH_t$ denote the $\sigma$-field generated by $\mcN(\tau), \tau \in [0,T]$. The intensity process $\blambda(t) = (\lambda_1(t),\ldots, \lambda_p(t))^\tr$ is a $p$-variate $\mcH_t$-predictable process defined as
$$\lambda_j(t)\d t = \mbP\big(\d N_j(t) = 1|\mcH_t\big),\ j \in [p].$$
In our model, we assume this intensity process takes the form
\begin{equation}
\label{EquationOriginalIntensity}
\lambda_j(t) = h_j\Big(\mu_j + \sum_{k=1}^p \int_0^t \omega_{j,k}(\tau) \d N_k(t - \tau)\Big),\ j \in [p],
\end{equation}
where $h_j(\cdot), j \in [p]$ are link functions and can be nonlinear. $\mu_j, j \in [p]$ are the background intensities of the components of the multivariate point process. $\omega_{j,k}(\cdot): \mbR^+ \rightarrow \mbR, j,k \in [p]  $ are the transfer functions. $\blambda(t) = \big(\lambda_1(t),\cdots,\lambda_p(t)\big)$ is called the intensity process. The convolutional expression $\int_0^t \omega_{j,k}(\tau) \d N_k(t - \tau)$ summaries the historical information of the $k$-th component. In this paper, we consider parameterized transfer functions that can be written as
$$\omega_{j,k}(t) = \beta_{j,k}\kappa_{j,k}(t),\ j,k \in [p],$$
where $\kappa_{j,k}(t),\ j,k \in [p]$ are known kernel functions.

In graphical models, the edges encodes conditional independence structures among the random variables (\citeauthor{maathuis2018handbook}, \citeyear{maathuis2018handbook}; \citeauthor{qiao2019functional}, \citeyear{qiao2019functional}; \citeauthor{engelke2020graphical}, \citeyear{engelke2020graphical}). For multivariate point processes, the question of interest is that whether the intensity of one component depends on the history of another component, which can be represent by the local independence for multivariate point process (\citeauthor{didelez2008graphical}, \citeyear{didelez2008graphical}).
\begin{define}[local independence for multivariate point process]
Let $\mcN = (N_1,\ldots,N_p)$ be a multivariate point process. Let further $A,B$ and $C$ be disjoint subsets of $[p]$. We then say that a subprocess $\mcN_B$ is locally independent of $\mcN_A$ given $\mcN_C$ over $\mcT$ if $\mbE[\lambda_k(t)|\mcF_t^{A\cup B\cup C}] = \mbE[\lambda_k(t)|\mcF_t^{B\cup C}]$ holds for all $t\in\ \mcT$ and $k \in B$, where $\mcF_t^A$ denotes the $\sigma$-field generated by $\{N_i([0,t]), i \in A\}$.
\end{define}

In temporal point process graphical models, the local independence structures are encoded in the transfer matrix $B = (\beta_{j,k}) \in \mbR^{p\times p}$, because $\lambda_j(t) = h_j\big(\mu_j + \sum_{k=1}^p \int_0^t \beta_{j,k}\kappa_{j,k}(\tau) \d N_k(t-\tau)\big) \in \mcF_t^A$ for index set $A$ as long as $\{k, \beta_{j,k} \neq 0\} \subset A$. We have the following proposition:
\begin{proposition}
\label{Propositionlocalindependence}
For a temporal point process graphical model $\mcN = (N_1,\ldots,N_p)$, let $A,B$ be disjoint subsets of $[p]$ and $\mcN_A = \{N_i,\ i \in A\}$. $\mcN_A$ is locally independent of $\mcN_B$ given $\mcN_{(A\cup B)^c}$ if and only if $\beta_{i,j} = 0,\ \forall i \in A,\ j \in B$.
\end{proposition}
Thus by estimating the transfer matrix $B$, we can recover the local independence relationships in $\mcN(t)$.

\section{Estimation}
In this section, we present an approach for estimating the temporal point process graphical models. We will begin with the construction of our estimator in Section \ref{SectionOurMethod}, then provide a detailed comparison among our method with the widely employed method in prior arts, the least-square estimation and the maximum likelihood estimation in Section \ref{SectionComparing}. In Section \ref{SectionTheoreticalGuarantees}, we will present theoretical analysis that guarantees the performance of our method.
\subsection{Our Method}
\label{SectionOurMethod}
In this paper, we will focus on the temporal point process graphical models with bounded link functions and compactly supported transfer functions, as explained in Assumption \ref{Assumption1}.
\begin{assumption}
\label{Assumption1}
There exists a constant $h_{\max}$ that $h_j(\cdot) < h_{\max}, j \in [p]$ and $h_j, j \in [p]$ have continuous derivatives. $\kappa_{j,k}(t), j,k \in [p]$ are supported on $[0,\kappa_{\supp}]$ for some constant $\kappa_{\supp}$ with bounded derivatives $|\kappa_{j,k}'(t)| \leq \kappa_{\max},j,k \in [p]$ (and naturally we can pick $\kappa_{\max}$ such that $|\kappa_{j,k}(t)| \leq \kappa_{\max}, j, k \in [p]$ also holds).
\end{assumption}
Similar versions of assumptions on the link function and transfer kernels are required in existing theory (\citeauthor{2015Hansen}, \citeyear{2015Hansen}; \citeauthor{chen2017nearly}, \citeyear{chen2017nearly}; \citeauthor{bacry2020sparse}, \citeyear{bacry2020sparse}). As we allow a nonlinear link function, the transfer kernels are no longer restricted to be non-negative. For simplicity of the theory, we assume that $h_j(\cdot)$ is bounded. This assumption covers most of the cases in the applications.

Under Assumption \ref{Assumption1}, the temporal point process is Markovian, i.e., given the event times in $[t,t+\kappa_{\supp}]$, the intensity process $\{\blambda(s), s > t + \kappa_{\supp}\}$ is independent of $\mcH_t$. Together with the fact that the intensity process is bounded by $h_{\max}$, it is straightforward to show that the empty state $\phi$, where there is no event on $[t,t+\kappa_{\supp}]$, is a positive recurrent and the temporal point process is stable in distribution. For more discussion of stability for the temporal point process, we refer interested readers to \citeauthor{Stability1996} (\citeyear{Stability1996}) and \citeauthor{2018Costa} (\citeyear{2018Costa}).

Suppose that we observe the event times of a stationary process on $[0,T]$. Under Assumption \ref{Assumption1}, we can rewrite the intensity in a compact form:
$$\lambda_j(t) = h_j\big(\mu_j + \langle \bx_j(t),\bbeta_j \rangle\big), j \in [p],$$
where $\bx_j(t) = \big(x_{j,1}(t),\cdots,x_{j,p}(t)\big)^\tr$, $x_{j,k}(t) = \int_0^t \kappa_{j,k}(t-\tau)\d N_k(\tau)$ and $\bbeta_j = (\beta_{j,1},\beta_{j,2},\ldots,\beta_{j,p})$. Inspired by the log-linear models in \citeauthor{negahban2009unified}(\citeyear{negahban2009unified}), we considered the following loss function for estimating $(\mu_j,\bbeta_j)$:
\begin{equation}
\label{EquationEmpiricalRiskMinimazation}
L_T(\bmu, B) = \frac{1}{T}\sum_{j=1}^{p}\int_0^T W_j(t)\big(H_j(\mu_j+\langle\bbeta_j,\bx_j(t)\rangle)\d t - \langle\bbeta_j,\bx_j(t)\rangle\d N_j(t)\big),
\end{equation}
where $H_j(\cdot)$ is a primitive function of $h_j(\cdot)$, i.e., $H_j'(\cdot) = h_j(\cdot)$, $\bmu = (\mu_1,\ldots,\mu_p)$ and $B = (\bbeta_1^\tr,\ldots,\bbeta_p^\tr)$, $\bW(t) = \big(W_1(t),\ldots, W_p(t)\big)$ is a weighting process.
If $\bmu$ and $B$ is intrinsically sparse due to some practical reasons, we can adopt the following $\ell_1$ penalized loss minimization (\citeauthor{tibshirani1996regression},\citeyear{tibshirani1996regression}):
\begin{equation}
\label{EquationPenalizedEmpiricalRiskMinimazation}
(\hat{\bmu},\hat{B}) \in \argmin_{\bmu,B} \Big\{\frac{1}{T}\sum_{j=1}^p\int_0^T W_j(t)\Big(H_j\big(\mu_j+\langle\bbeta_j,\bx_j(t)\rangle\big)\d t - \langle\bbeta_j,\bx_j(t)\rangle\d N_j(t)\Big) + \lambda_T (\|\bmu\|_1 + \|B\|_1)\Big\},\ j \in [p].
\end{equation}

In the generalized linear models, \citeauthor{negahban2009unified}(\citeyear{negahban2009unified}) employed a similar method to derive a convex loss function. The negative log-likelihood loss function for generalized linear model is convex but not strongly convex when $p > T$, i.e., there are some directions that have zero gradient and the Hessian matrix is not of full rank. \citeauthor{negahban2009unified} (\citeyear{negahban2009unified}) addressed this problem by introducing the conception of restricted strong convexity. For a decomposable penalty (such as $\|\cdot\|_1$) and a convex loss function, the solutions of optimization will always fall in a 'restricted set', restricted to which the loss function is strongly convex. We show that our loss function also admits a restriction strong convexity condition. This implies that the obtained estimator, regardless of algorithm used or initial value, will provide the convergence rate we proved. In contrast, the estimator provided by \citeauthor{tang2021multivariate} (\citeyear{tang2021multivariate}) is a local estimator, which only retain the statistical property as long as the initial value is close enough to the true parameters.

\subsection{Examples}
\label{SectionComparing}
For estimating a temporal point process graphical with linear link functions, there are two widely used estimating strategies, the least-square estimator (\citeauthor{bacry2020sparse},\citeyear{bacry2020sparse}; \citeauthor{2015Hansen},\citeyear{2015Hansen}):
\begin{equation}
\label{EquationLSRisk}
L^{\mathrm{LS}}_T(\bmu,B) = \frac{1}{T}\sum_{j=1}^p \int_0^T \lambda_{j}^2 (t; \mu_j,\bbeta_j) \d t - \frac{2}{T}\int_0^T \lambda_{j}(t; \mu_j,\bbeta_j) \d N_j(t),
\end{equation}
and the maximum likelihood estimator (\citeauthor{ozaki1979maximum},\citeyear{ozaki1979maximum}; \citeauthor{tang2021multivariate}, \citeyear{tang2021multivariate})
\begin{equation}
\label{EquationMLERisk}L^{\mathrm{MLE}}_{T}(\bmu,B) = \frac{1}{T}\sum_{j=1}^p\int_0^T \lambda_j (t;\mu_j,\bbeta_j)\d t - \frac{1}{T}\int_0^T \log \big(\lambda_j(t; \mu_j,\bbeta_j)\big)\d N_j(t).
\end{equation}

The score functions for least square loss and negative log-likelihood loss are

$$\nabla_{\bbeta_j} L_T^{\mathrm{LS}} = \frac{2}{T}\int_0^T h'_j(\mu_j + \langle\bbeta_j ,\bx_j(t)\rangle)\bx_j(t) \Big[h_j(\mu_j + \langle\bbeta_j ,\bx_j(t)\rangle)  \d t - \d N_j(t) \Big]$$
and
$$\nabla_{\bbeta_j} L_T^{\mathrm{MLE}} = \frac{1}{T}\int_0^T \frac{h'_j(\mu_j + \langle\bbeta_j ,\bx_j(t)\rangle)}{h_j(\mu_j + \langle\bbeta_j ,\bx_j(t)\rangle)}\bx_j(t) \Big[h_j(\mu_j + \langle\bbeta_j ,\bx_j(t)\rangle)  \d t - \d N_j(t) \Big].$$

It is seen that optimizing the least-square risk function is equivalent to optimizing \eqref{EquationEmpiricalRiskMinimazation} with iterative reweigh $\hat{W}^{\mathrm{LS}}_j(t) = h'_j(\mu_j + \langle\hat{\bbeta}_j,\bx_j(t)\rangle)$ and optimizing the negative log-likelihood risk function is equivalent to optimizing \eqref{EquationEmpiricalRiskMinimazation} with iterative reweigh $\hat{W}^{\mathrm{MLE}}_j(t) = \frac{h'_j(\mu_j + \langle\hat{\bbeta}_j,\bx_j(t)\rangle)}{h_j(\mu_j + \langle\hat{\bbeta}_j,\bx_j(t)\rangle)}$. Directly optimizing the risk function for least square estimator and maximum likelihood estimator are non-convex problems. But our iterative reweigh procedure permits us to approximate the non-convex problem with a sequence of convex problems.

\subsection{Theoretical Guarantees}
\label{SectionTheoreticalGuarantees}
Now we return to the $\ell_1$-penalized loss minimization \eqref{EquationPenalizedEmpiricalRiskMinimazation}. Theoretical result indicates that our estimator is consistent even in high-dimensional case. When unrestricted, it is possible to cook up extreme graphs, where, for instance, some components of the influence process $\bx(t)$ are constantly pushed towards infinity, which makes the model hardly identifiable. To avoid such cases, we pose the following regularity conditions.

\begin{assumption}
\label{Assumption2}

(1)The mean influence, $\mu^*_j + \sum_{k \in S_j} \beta^*_{j,k}\mbE[x_{j,k}]$ are uniformly bounded:
$$\sup_{j \in [p]} \mu^*_j + \sum_{k \in S_j} \beta^*_{j,k} \mbE[x_{j,k}] < \sigma_0,\ \forall j \in [p],$$
for some constant $\sigma_0 > 0$ for every $j$.

(2) The covariance matrix of the influence process $\Cov\big(\bx_j(t)\big) =\Cov\big(x_{j,1}(t),\cdots,x_{j,p}(t)\big) = \Sigma_j$ is positive definite and has bounded eigenvalue:
$$0 < \Sigma_{\min} \leq \Gamma_{\min}(\Sigma_j) \leq \Gamma_{\max}(\Sigma_j) \leq \Sigma_{\max} < \infty,\ \forall j \in [p]$$
where $\Gamma_{\min}(\cdot)$ and $\Gamma_{\max}(\cdot)$ indicate the smallest and the largest eigenvalue of a matrix. That is, for any $(\delta_\mu,\bDelta^\tr) \in \mbS^{p+1}_2(1)$ we have
$$\mbE \big(\delta_\mu + \langle\bDelta, \bx_j(t)\rangle\big)^2 > \sigma_1 $$
and there exists some $R > 0$ such that
\begin{equation}
\label{EquationStrongerLambda}
\mbE \big(\delta_\mu + \langle\bDelta, \bx_j(t)\rangle\big)^2\cdot \mbI\big(|\mu_i^* + \langle\bbeta_j^*,\bx_j(t)\rangle| \leq R\big) \cdot \mbI\big(|\delta_\mu + \langle\bDelta,\bx_j(t)\rangle| \leq R\big) > \sigma_1.
\end{equation}

(3) The derivative is bounded away from $0$ on a compact interval: there exists $\sigma_2 > 0$ such that
$$\inf_{-2R < t < 2R} h_j'(t) \geq \sigma_2 > 0,\ \forall j \in [p],$$
where $R$ is introduced above.

(4) There exist $0 < \sigma_3 < \sigma_4$ that $ \sigma_3 < W_j(t) <\sigma_4,\ \forall j \in [p]$ with probability $1$ for $T$ large enough.
\end{assumption}

We briefly discuss these assumptions. Assumption \ref{Assumption2} (1) restricts the average influence of process $N_j(t)$ and it is equivalent to having bounded intercept terms in vanilla Lasso problems (\citeauthor{tibshirani1996regression}, \citeyear{tibshirani1996regression}).
Having Non-degenerate covariates (Assumption \ref{Assumption2} (2)) is a standard requirement for estimation problems (condition GLM1 in \citeauthor{negahban2009unified},\citeyear{negahban2009unified}, regularity condition 4 in \citeauthor{tang2021multivariate}, \citeyear{tang2021multivariate}). Here \eqref{EquationStrongerLambda} asks for a version of non-degenerate that looks stronger. It is not truly stronger. If we let $R \rightarrow \infty$, we will have
$$\mbE \big(\delta_\mu + \langle\bDelta, \bx_j(t)\rangle\big)^2\cdot \mbI\big(|\mu_i^* + \langle\bbeta_j^*,\bx_j(t)\rangle| \leq R\big) \cdot \mbI\big(|\delta_\mu + \langle\bDelta,\bx_j(t)\rangle| \leq R\big) \rightarrow \mbE \big(\delta_\mu + \langle\bDelta, \bx_j(t)\rangle\big)^2.$$
This involves a trade-off between constants $\sigma_1$ and $\sigma_2$. One can show that \eqref{EquationStrongerLambda} holds if $\bx_j(t)$ has a sub-exponential tail.
Assumption \ref{Assumption2} (3) is employed to ensure identifiability. To conclude, Assumption \ref{Assumption2} (2) and \ref{Assumption2} (3) are required mainly for technical reasons. We propose Assumption \ref{Assumption2} (4) to make sure that the weights are well-behaved.

Now we are ready to introduce the main theorem.

\begin{theorem}
\label{TheoremMainTheorem}
For estimators in \eqref{EquationPenalizedEmpiricalRiskMinimazation}, under Assumption \ref{Assumption1} and \ref{Assumption2}, there exists constants $c_1,c_2,c_3> 0$ that with $\lambda_T = c_1 \sqrt{\frac{\tau\log(p\vee T)}{T}}$,
we have
\begin{equation}
\label{EquationL2JBound}
p^{-1/2}\|(\hat{\bmu},\hat{B}) - (\bmu^*,B^{*})\|_F \leq c_2 \sqrt{s}\lambda_T
\end{equation}

\begin{equation}
\label{EquationL1JBound}\|(\hat{\bmu},\hat{B}) - (\bmu^*,B^{*})\|_\infty \leq c_2 s\lambda_T\end{equation}
hold with possibility greater than $1 - \frac{c_3}{(p\vee T)^{\tau}}$, where $s = 1 + \max_j \{\|\bbeta_{j}\|_0, j \in [p]\}$ and a star $*$ as a superscript denotes the true value of parameters.
\end{theorem}
Theorem \ref{TheoremMainTheorem} guarantees that with high probability, the estimator will converge in the order of $\sqrt{\frac{s\log(p \vee T)}{T}}$, which is similar to the established ones in prior works (\citeauthor{bacry2020sparse}, \citeyear{bacry2020sparse}; \citeauthor{2015Hansen}, \citeyear{2015Hansen}; \citeauthor{tang2021multivariate}, \citeyear{tang2021multivariate}). It is important to note that Theorem \ref{TheoremMainTheorem} considers an \textit{oracle} procedure, where the tuning parameters depend on unknown parameters. When the linear functions are linear, \citeauthor{2015Hansen} (\citeyear{2015Hansen}) explored the selection guidelines in the case that the number of nodes $p$ is fixed but the dictionary for sieving transfer kernels is growing. A general theory for tuning parameter selection when $p \rightarrow \infty$ still remains an open problem. A detailed proof of Theorem \ref{TheoremMainTheorem} is provided in the Appendix.

\section{Implementation}
\subsection{Theory for Approximation}
Implementation lies between the beautiful statistical theory and the successful application of statistical methods. The concrete details of implementation are important but seldom mentioned in prior articles. In this section, we present a comprehensive analysis of the implementative version of our estimation.

For temporal point processes, the algorithm usually proceeds in a discrete manner (\citeauthor{tang2021multivariate}, \citeyear{tang2021multivariate}). The observation interval $[0,T]$ is equally divided into $M $ subintervals and the length of each interval is $\frac{T}{M}$ and the integral is replaced by numerical integral. Let $y_{j,m} = N_j(\frac{(m+1)T}{M}) - N_j(\frac{mT}{M})$ denote the number of events in $[\frac{mT}{M}, \frac{(m+1)T}{M}]$ of node $j$ and let $x_{j,k,m} = x_{j,k}(\frac{mT}{M})$ for $m = 0,\ldots, M-1$. Let $\bx_{j,m} = (\bx_{j,1,m},\cdots,\bx_{j,p,m})$. Then our loss function in implementation becomes
\begin{equation}
\label{EquationEmpiricalRiskMinimazationD}\tilde{L}_T(\bmu,B) = \frac{1}{T}\sum_{j=1}^p\sum_{m=0}^{M-1}W_j(\frac{mT}{M})\Big(\frac{T}{M}H_j(\mu_j + \langle\bbeta_j,\bx_{j,m}\rangle) - y_{j,m}(\mu_j+\langle\bbeta_j,\bx_{j,m}\rangle)\Big).
\end{equation}

Generally, the discrete version \eqref{EquationEmpiricalRiskMinimazationD} has additional approximation error compared to \eqref{EquationEmpiricalRiskMinimazation}. To  characterize the additional error, we pose the following error decomposition:
$$y_{j,m} = N_j(\frac{(m+1)T}{M}) - N_j(\frac{mT}{M}) = \mbE[N_j(\frac{(m+1)T}{M}) - N_j(\frac{mT}{M}) \Big| \lambda_j|_{(0,\frac{(m+1)T}{M}]}] + \epsilon_{1,j,m}$$
and
\begin{align*}
\mbE[N_j(\frac{(m+1)T}{M}) - N_j(\frac{mT}{M}) \Big| \lambda_j|_{(0,\frac{(m+1)T}{M}]}] + \epsilon_{1,j,m} &= \int_{\frac{mT}{M}}^{\frac{(m+1)T}{M}}h_j(\mu_j^* + \langle\bx_j(\tau),\bbeta_j^*\rangle) \mathrm{d} \tau\\
& = \frac{T}{M}h_j(\mu_j^*+\langle\bx_{j,m},\bbeta^*_j\rangle) + \epsilon_{2,j,m},
\end{align*}
where $\bx_{j,m} = (x_{j,1,m}, x_{j,2,m}, \ldots,x_{j,p,m})$. Thus we have
$$y_{j,m} = \frac{T}{M}h_j\Big(\mu_j^* + \langle\bx_{j,m},\bbeta_j^*\rangle\Big) + \epsilon_{j,m},\ j \in [p],\ m = 0,\ldots, M-1,$$
where $\epsilon_{j,m} = \epsilon_{1,j,m} + \epsilon_{2,j,m}$. The choice of $M$ is of great importance because it leads a trade-off between computation efficiency and approximation accuracy. To ensure the approximation error $\epsilon_{2,j,m}$ is dominated by the statistical noise $\epsilon_{1,j,m}$, we assume that,
\begin{assumption}
\label{Assumption3}
 $M$, the number of subintervals is large enough that $\frac{\tau^{1/4}s^{1/2}T^{5/4}\log^2(p\vee T)}{M} = o(1)$.
\end{assumption}
The $\ell_1$ penalized loss minimization in implementation is
\begin{equation}
\label{EquationPenalizedEmpiricalRiskMinimazationD}(\tilde{\bmu},\tilde{B}) \in \argmin_{\bmu,B} \Big\{\frac{1}{T}\sum_{j=1}^p \sum_{m=0}^{M-1}W_j(\frac{mT}{M})\Big(\frac{T}{M}H_j(\mu_j + \langle\bbeta_j,\bx_{j,m}\rangle) - y_{j,m}(\mu_j+\langle\bbeta_j,\bx_{j,m}\rangle)\Big) + \lambda_T(\|\bmu\|_1 + \|B\|_1)\Big\}.
\end{equation}

By carefully analysing the approximation error, we can provide a similar bound for convergence rate for $(\tilde{\bmu},\tilde{B})$:
\begin{theorem}
For estimators in \eqref{EquationPenalizedEmpiricalRiskMinimazationD}, under Assumption \ref{Assumption1}-\ref{Assumption3}, we have there exists constants $c_1,c_2,c_3> 0$ that with $\lambda_T = c_1 \sqrt{\frac{\tau\log(p\vee T)}{T}}$,
we have
\begin{equation}
\label{EquationL2JBoundD}
p^{-1/2}\|(\tilde{\bmu},\tilde{B}) - (\bmu^*,B^{*})\|_F \leq c_2 \sqrt{s}\lambda_T
\end{equation}

\begin{equation}
\label{EquationL1JBoundD}\|(\tilde{\bmu},\tilde{B}) - (\bmu^*,B^{*})\|_\infty \leq c_2 s\lambda_T\end{equation}
hold with possibility greater than $1 - \frac{c_3}{(p\vee T)^{\tau}}$, where $s = 1 + \max_j \{\|\bbeta_{j}\|_0, j \in [p]\}$ and a star $*$ as a superscript denotes the true value of parameters.
\end{theorem}
When $M$ satisfies Assumption \ref{Assumption3}, the approximation error is dominated by the statistical noise. There are other graphical models of point process that divided the observation window into a number of bins and model the number of events in each bin, with a generalized linear model (\citeauthor{Chunming2016},\citeyear{Chunming2016}) or with Gaussian graphical model (\citeauthor{vinci2018adjusted}, \citeyear{vinci2018adjusted}). Our method is fundamentally different from these methods because discretization is employed only in the implementation of the method. Our final goal is to estimate the parameters in the generalized Hawkes process instead of other models. A detailed proof is included in the supplementary material.

\section{Simulation}
\label{SectionSimulationStudy}
In this section, we investigate the performance of our estimation procedure in a simulation study. We propose two estimators, the naively weighted estimator with $W_j(t) = 1,\ \forall j \in [p]$ and the iterative reweighted maximum likelihood estimator where $W_j(t) = \frac{h_j'(t)}{h_j(t)}$ is iteratively updated.

\begin{figure}
  \centering
  % Requires \usepackage{graphicx}
  \includegraphics[width=15cm]{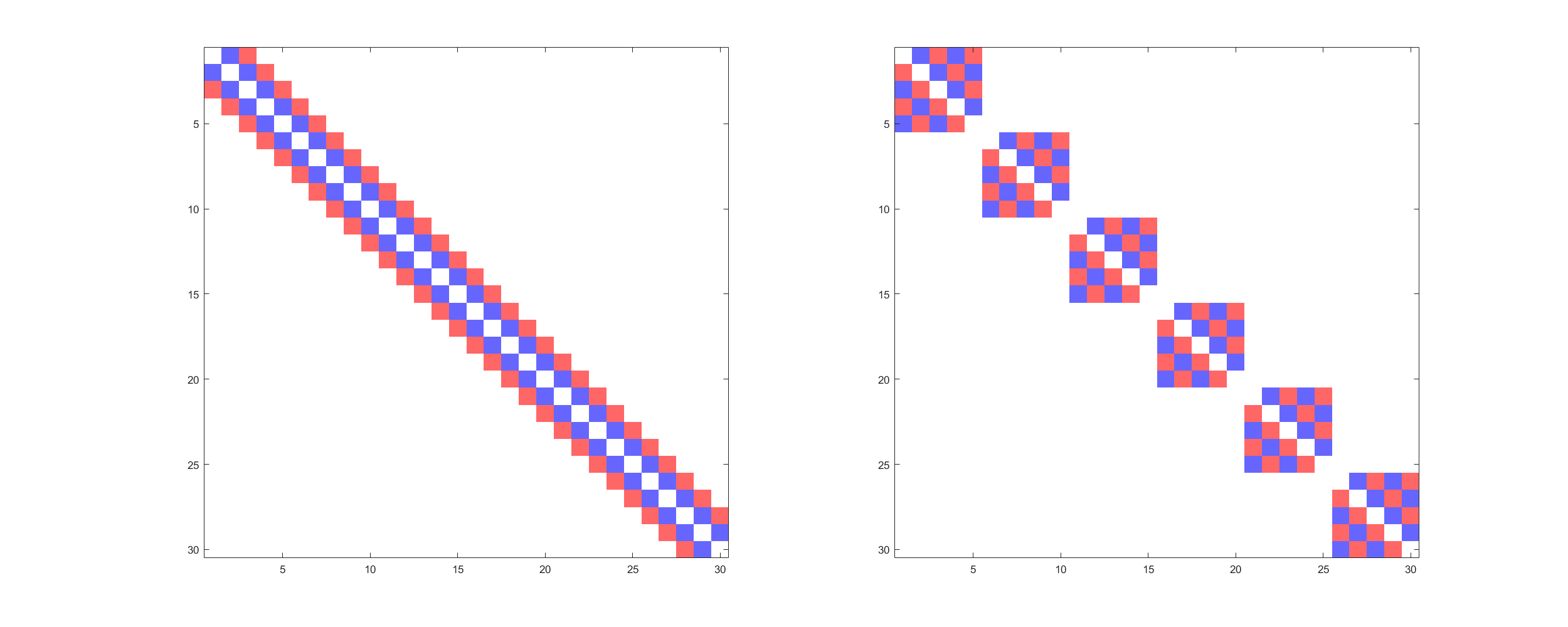}\\
  \caption{Connectivity matrices with block and chain structures for $p = 30$. The excitatory connection are shown in blue with entry $\beta_{j,k} = 0.3$ and the inhibitory connection are shown in red with entry $\beta_{j,k} = -0.3$. }\label{Figure1Structure}
\end{figure}

\begin{figure}
  \centering
  % Requires \usepackage{graphicx}
  \includegraphics[width=15cm]{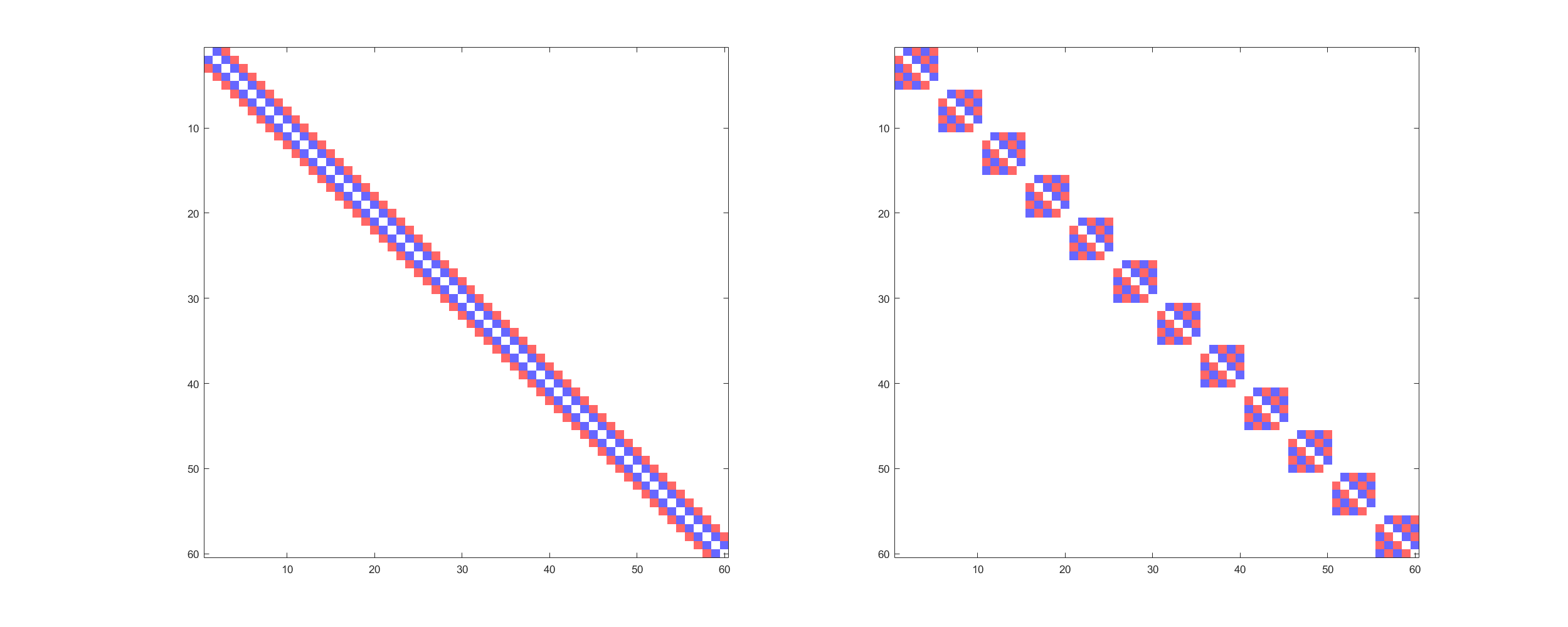}\\
  \caption{Connectivity matrices with block and chain structures for $p = 60$. The excitatory connection are shown in blue with entry $\beta_{j,k} = 0.3$ and the inhibitory connection are shown in red with entry $\beta_{j,k} = -0.3$. }\label{Figure2Structure}
\end{figure}

We set the number of nodes to $p \in \{30,60\}$. We consider two popular graphical structures: block and chain.
For the block structure, $B$ is a $60\time 60$ ($30 \times 30$) blockwise diagonal matrix of 12 (6) identical symmetrical Toeplitz matrices whose first row is $(0,0.3,-0.3,0.3,-0.3)$ for $p = 60$ ($p = 30$).
For the chain structure, $B$ is a symmetrical Toeplitz matrix whose first row is $(0,0.3,-0.3,0,0,\cdots,0)$.
Their explicit forms are shown in Figure \ref{Figure1Structure} and \ref{Figure2Structure} for $p=30$ and $p=60$, where the excitatory connection are shown in blue with entry $\beta_{j,k} = 0.3$ and the inhibitory connection are shown in red with entry $\beta_{j,k} = -0.3$.
The background intensity is $\mu_j = 0.5$. We consider two sets of link function and transfer kernel. For setting 1, we consider an $\arctan$ function as the link function and a restricted linear transfer kernel.

\begin{align*}
\begin{cases}
h_j(x) & = 4 + \frac{8}{\pi} \arctan(x),\ j \in [p], \\
\kappa_{j,k}(x) & = (1 - x)\mbI_{[0,1]}(x),\ j,k \in [p].
\end{cases}
\end{align*}
 For setting 2, we consider a sigmoid function as our link function and a exponential transfer kernel.

\begin{align*}
\begin{cases}
h_j(x) & = 1 + \frac{4 e^x}{1 + e^x},\ j \in [p], \\
\kappa_{j,k}(x) & = e^{-x}\mbI_{[0,5]}(x),\ j,k \in [p].
\end{cases}
\end{align*}

The length of observation is $T \in \{200,400\}$ and the number of sub-intervals $M = 10T$. We use a 5-fold cross-validation to select the penalty level. We evaluate the estimation accuracy by the relative $\ell_1$ error $\frac{\sum_{i,j}|\hat{\beta}_{i,j} -\beta_{i,j} |}{\sum_{i,j}|\beta_{i,j}|}$ and the relative mean square error of the estimated transferring coefficient matrix $\frac{\|\hat{B} - B^*\|_F^2}{\|B^*\|_F^2}$. Three methods are compared: the $\ell_1$ penalized naively weighted estimator (Sparse-Naive) the $\ell_1$ penalized iteratively reweighted maximum likelihood estimator (Sparse-MLE) and the vanilla maximum likelihood estimator. The average relative $\ell_1$ error and $\ell_2$ error based on 50 replications with the standard errors in the parenthesis are reported.

\begin{table}
\centering
\caption{Estimation accuracy of $B$ in form of relative $\ell_1$ error}
\label{TableMetric1}
\begin{tabular}{ccccccc}
\hline
Setting    & Structure& p  &  T  & Sparse-Naive  &   Sparse-MLE &   vanilla-MLE \\
\hline
Setting 1  &   Block  & 30 & 200 & 0.718(0.049) & \textbf{0.667(0.025)} & 1.510(0.057)\\
           &          &    & 400 & \textbf{0.525(0.024)} & 0.547(0.032) & 0.942(0.025)\\
           &          & 60 & 200 & 1.010(0.077) & \textbf{0.805(0.022)} & 4.460(0.134)\\
           &          &    & 400 & 0.776(0.037) & \textbf{0.656(0.011)} & 2.128(0.040)\\
           &  Chain   & 30 & 200 & 0.877(0.043) & \textbf{0.796(0.032)} & 1.793(0.135)\\
           &          &    & 400 & \textbf{0.645(0.028)} & 0.656(0.034) & 1.066(0.048)\\
           &          & 60 & 200 & 1.240(0.086) & \textbf{0.959(0.025)} & 5.277(0.170)\\
           &          &    & 400 & 0.916(0.049) & \textbf{0.787(0.022)} & 2.403(0.063)\\
Setting 2  &   Block  & 30 & 200 & 0.895(0.037) & \textbf{0.884(0.033)} & 2.812(0.163)\\
           &          &    & 400 & \textbf{0.716(0.023)} & 0.723(0.026) & 1.612(0.052)\\
           &          & 60 & 200 & \textbf{1.035(0.043)} & 1.048(0.032) & 14.62(0.611)\\
           &          &    & 400 & \textbf{0.862(0.025)} & 0.872(0.025) & 3.906(0.083)\\
           &  Chain   & 30 & 200 & 1.063(0.057) & \textbf{1.054(0.033)} & 3.571(0.213)\\
           &          &    & 400 & 0.883(0.037) & \textbf{0.882(0.038)} & 1.826(0.086)\\
           &          & 60 & 200 & \textbf{1.174(0.044)} & 1.176(0.026) & 15.77(0.546)\\
           &          &    & 400 & 1.031(0.029) & \textbf{1.024(0.024)} & 4.451(0.137)\\
\hline
\end{tabular}
			%\begin{tablenotes}
            %\item Notes: Three methods are compared: the $\ell_1$ penalized naively weighted estimator (Sparse-Naive) the $\ell_1$ penalized iterative reweighted maximum likelihood estimator (Sparse-MLE) and the vanilla maximum likelihood estimator. Reported are the average relative $\ell_1$ error based on 50 replications, with the standard errors in the parenthesis.
            %\end{tablenotes}

\end{table}

\begin{table}
\centering
\caption{Estimation accuracy of $B$ in form of relative $\ell_2$ error}
\label{TableMetric2}
		%\centering
\begin{tabular}{ccccccc}
\hline
Setting    & Structure& p  &  T  & Sparse-Naive  &   Sparse-MLE &   vanilla-MLE \\
\hline
Setting 1  &   Block  & 30 & 200 & \textbf{0.235(0.018)} & 0.241(0.018) & 0.495(0.042)\\
           &          &    & 400 & \textbf{0.149(0.012)} & 0.150(0.012) & 0.191(0.011)\\
           &          & 60 & 200 & \textbf{0.300(0.015)} & 0.307(0.014) & 2.235(0.143)\\
           &          &    & 400 & \textbf{0.174(0.005)} & 0.194(0.007) & 0.483(0.019)\\
           &  Chain   & 30 & 200 & 0.326(0.029) & \textbf{0.318(0.026)} & 0.843(0.301)\\
           &          &    & 400 & 0.200(0.017) & \textbf{0.181(0.018)} & 0.250(0.034)\\
           &          & 60 & 200 & 0.418(0.027) & \textbf{0.399(0.021)} & 3.523(0.296)\\
           &          &    & 400 & \textbf{0.224(0.011)} & 0.242(0.011) & 0.687(0.077)\\
Setting 2  &   Block  & 30 & 200 & 0.485(0.030) & \textbf{0.475(0.032)} & 1.772(0.268)\\
           &          &    & 400 & 0.285(0.020) & \textbf{0.281(0.020)} & 0.555(0.036)\\
           &          & 60 & 200 & 0.575(0.029) & \textbf{0.566(0.024)} & 24.53(1.682)\\
           &          &    & 400 & 0.348(0.020) & \textbf{0.341(0.019)} & 1.652(0.074)\\
           &  Chain   & 30 & 200 & 0.672(0.083) & \textbf{0.620(0.045)} & 3.575(0.729)\\
           &          &    & 400 & 0.389(0.033) & \textbf{0.371(0.027)} & 0.736(0.120)\\
           &          & 60 & 200 & 0.769(0.064) & \textbf{0.724(0.058)} & 27.85(1.561)\\
           &          &    & 400 & 0.484(0.023) & \textbf{0.462(0.022)} & 2.561(0.266)\\
\hline
\end{tabular}
			%\begin{tablenotes}
            %\item Notes: Three methods are compared: the $\ell_1$ penalized naively weighted estimator (Sparse-Naive) the $\ell_1$ penalized iterative reweighted maximum likelihood estimator (Sparse-MLE) and the vanilla maximum likelihood estimator. Reported are the average relative $\ell_1$ error based on 50 replications, with the standard errors in the parenthesis.
            %\end{tablenotes}

\end{table}

\begin{table}
\centering
\caption{The mean area under the ROC curves}
\label{TableMetricAUC}
		%\centering
\begin{tabular}{cccccc}
\hline
Setting    & Structure& p  &  T  & Sparse-Naive  &   Sparse-MLE \\
\hline
Setting 1  &   Block  & 30 & 200 & 0.995(0.0026) & \textbf{0.997(0.0015)} \\
           &          &    & 400 & \textbf{1.000(0.0001)} &\textbf{ 1.000(0.0001)} \\
           &          & 60 & 200 & 0.995(0.0016) & \textbf{0.997(0.0014)} \\
           &          &    & 400 & \textbf{1.000(0.0001)} & \textbf{1.000(0.0001)} \\
           &  Chain   & 30 & 200 & 0.954(0.0074) & \textbf{0.965(0.0063)} \\
           &          &    & 400 & 0.974(0.0025) & \textbf{0.978(0.0027)} \\
           &          & 60 & 200 & 0.968(0.0041) & \textbf{0.976(0.0032)} \\
           &          &    & 400 & 0.985(0.0012) & \textbf{0.988(0.0012)} \\
Setting 2  &   Block  & 30 & 200 & 0.914(0.0143) & \textbf{0.916(0.0139)} \\
           &          &    & 400 & 0.979(0.0044) & \textbf{0.981(0.0042)} \\
           &          & 60 & 200 & 0.915(0.0137) & \textbf{0.918(0.0138)} \\
           &          &    & 400 & 0.983(0.0039) & \textbf{0.985(0.0035)} \\
           &  Chain   & 30 & 200 & 0.837(0.0174) & \textbf{0.841(0.0189)} \\
           &          &    & 400 & 0.924(0.0090) & \textbf{0.925(0.0100)} \\
           &          & 60 & 200 & 0.840(0.0134) & \textbf{0.844(0.0145)} \\
           &          &    & 400 & 0.944(0.0056) & \textbf{0.948(0.0052)} \\
\hline
\end{tabular}
			%\begin{tablenotes}
            %\item Notes: Three methods are compared: the $\ell_1$ penalized naively weighted estimator (Sparse-Naive) the $\ell_1$ penalized iterative reweighted maximum likelihood estimator (Sparse-MLE) and the vanilla maximum likelihood estimator. Reported are the average relative $\ell_1$ error based on 50 replications, with the standard errors in the parenthesis.
            %\end{tablenotes}
\end{table}

Table \ref{TableMetric1} and \ref{TableMetric2} summarize the results based on $50$ data replications. It is seen that both of our proposed methods outperform the vanilla maximum likelihood estimator, which is known to be asymptotically efficient (\citeauthor{ogata1978asymptotic},\citeyear{ogata1978asymptotic}) in estimating parameters when the transfer function is linear. Overall, the iteratively reweighted penalized maximum likelihood estimator achieves better performance. But the loss surface for the $\ell_1$ penalized MLE is not convex, so there is no permission that the working algorithm will find a feasible solution. The naive weighted estimator provides a convex loss function that guarantees that the algorithm returns a solution in the neighborhood of a feasible solution.

\begin{figure}[htbp]
\begin{minipage}[t]{0.5\textwidth}
\includegraphics[width = 7cm]{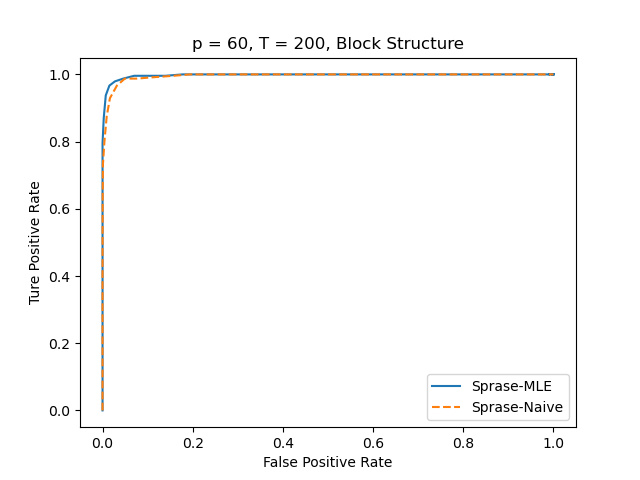}
\end{minipage}
\begin{minipage}[t]{0.5\textwidth}
\includegraphics[width = 7cm]{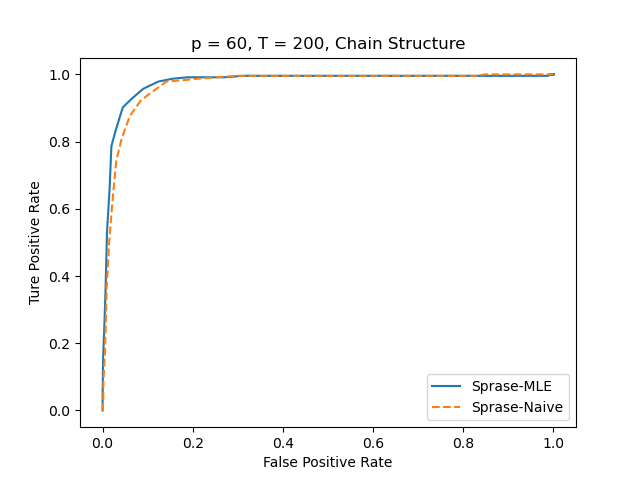}
\end{minipage}

\begin{minipage}[t]{0.5\textwidth}
\includegraphics[width = 7cm]{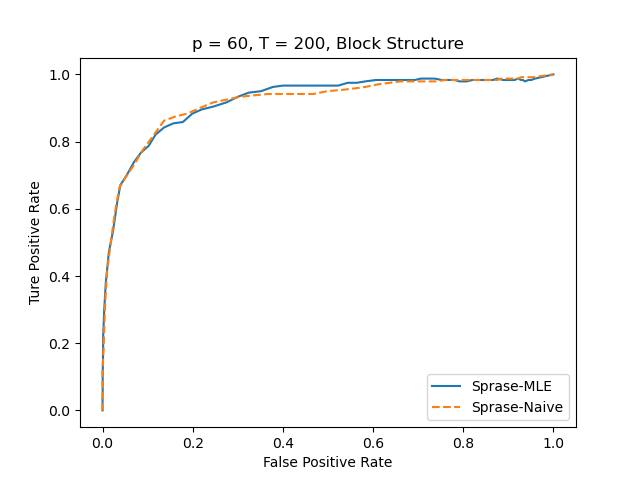}
\end{minipage}
\begin{minipage}[t]{0.5\textwidth}
\includegraphics[width = 7cm]{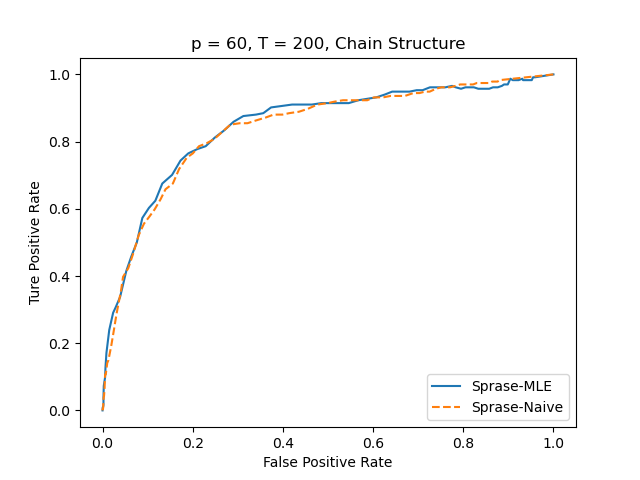}
\end{minipage}
\caption{ROC curves for $p=60, T = 200$ in setting 1 (top row) and setting 2 (bottom row).}
\label{FigureROC}
\end{figure}

To demonstrate the accuracy of variable selection for our proposed methods, we calculated true-positive rates and false-positive rates, i.e, the edges of the network that were correctly identified for each method and tuning parameter. Plotting true positive rate versus false positive rate over a fine grid of values of the tuning parameter produces a ROC curve, with curves near the top left corner indicating a method that performs well. Figure \ref{FigureROC} shows the median of the ROC curves for our two proposed methods and table \ref{TableMetricAUC} provides the area under the ROC curve (average over the 50 simulation runs). It is seen that both methods are successful in recovering the graphical structure in all the settings we considered.

Finally, we report the computation time. The naively weighted estimator has a convex loss function and does not need to calculate the reweigh parameters and runs faster than the iteratively reweigh MLE. The simulation code is programmed in Python. For the simulation example in setting 1 with $p = 60$, $T = 400$ and $M = 4000$, the average computing time was about 1.5 minutes for Naive estimator and 1.7 minutes for MLE. For the simulation example in setting 2 with $p = 60$, $T = 400$ and $M = 4000$, the average computing time was $2.1$ minutes for Naive estimator and $2.5$ minutes for MLE. All computations were performed with a Intel(R) Core(TM) i9 10900K CPU@3.7GHz.
\section{Application}
We consider the task of recovering the connectivity network among the neurons using the spike train data from the optogenetic study of \citeauthor{2018Rec}(\citeyear{2018Rec}) discussed in the Section \ref{SectionIntroduction}. In this experiment, the spike trains are recorded at 30 kHz on the olfactory bulb (OB) and the piriform cortex (PCx) of the mice, while a laser pulse is applied directly on the OB cells. For specially designed transgenic mice, light pulses will elicited an increase in OB spiking and remain elevated for the duration of the stimulus. We consider the spike train data collected at three intensity levels, $0\ mW/mm^2$, $5\ mW/mm^2$  and $10\ mW/mm^2$ from $15$ OB cells and $45$ PCx cells. For each intensity level, the observation lasts about $T = 90$ seconds. We choose the number of sub-intervals $M = 2000$ by pre-experiments.

\begin{figure}[htbp]
\begin{minipage}[t]{\textwidth}
\includegraphics[width = 17cm]{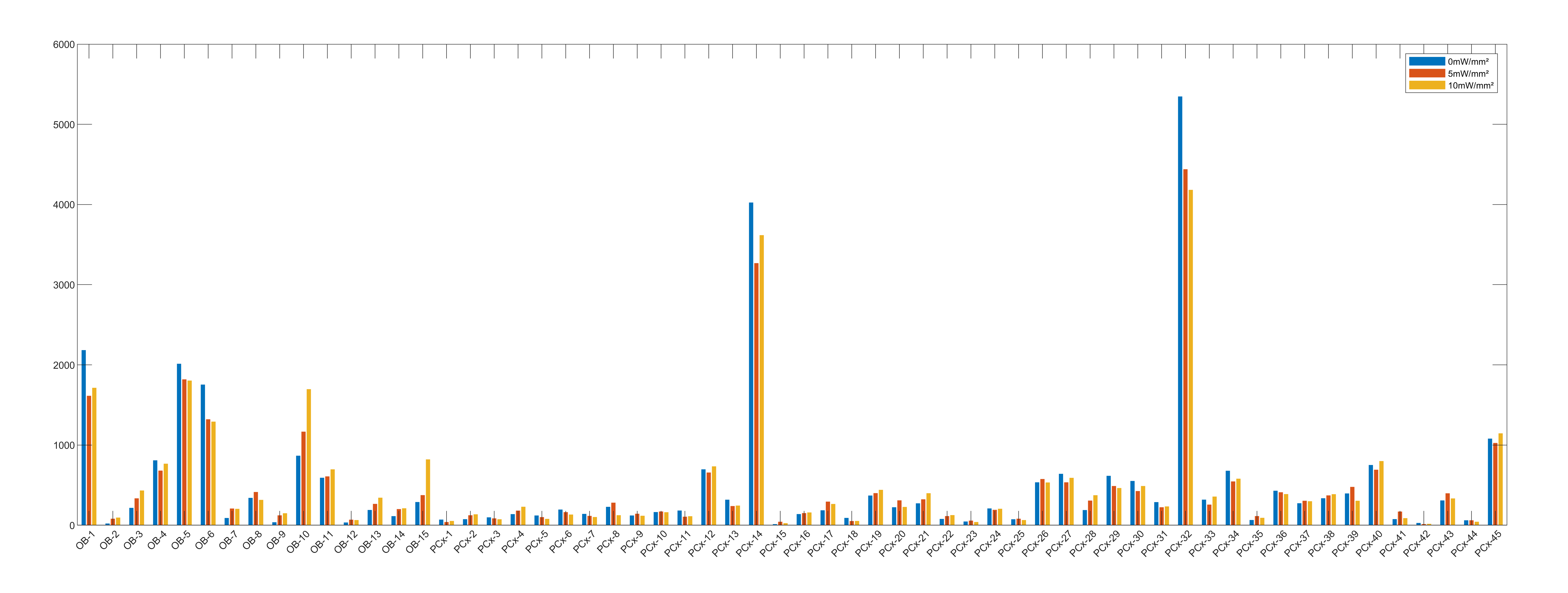}
\end{minipage}

\caption{Number of firings of the OB cells and the PCx cells for laser intensity $0\ mW/mm^2$, $5\ mW/mm^2$ and $10\ mW/mm^2$. OB cells are numbered 1-15 and PCx cells are numbered 16-60.}
\label{FigureApplicationNumberOfPoints}
\end{figure}

Figure \ref{FigureApplicationNumberOfPoints} shows the number of firings of OB cells and PCx cells. The firing numbers vary largely among cells. We use an adaptive link-function to better characterize the average intensity of different neurons:
$$h_i(t) = \hat{\lambda}_i \big(1 + \frac{2\arctan(t)}{\pi}\big),\ i = 1,\ldots, 60,$$
where $\hat{\lambda}_i = \frac{N_i([0,T])}{T}$. Based on the research of the duration of influence among the neurons in \citeauthor{2018Rec}(\citeyear{2018Rec}), we choose $\kappa_{j,k}(x) = \mbI_{[0,0.25]}(x), j,k, \in [p]$. Since our goal was to provide interpretable visualizations and investigate the influence of the laser on the neural connectivity, we computed sparse graphs with approximately $5\%$ connected edges. We performed a bootstrap procedure by randomly selecting $2000$ sub-intervals with replacement from the original data, finding a tuning parameters $\gamma_n$ to achieve $5\%$ sparsity level, fitting the point process graphical model and repeating 50 times. The final graph was constructed from the excitatory and inhibitory edges that occurred in at least $50\%$ of the bootstrap replications.
\begin{figure}[htbp]
\begin{minipage}[t]{0.30\textwidth}
\includegraphics[width = 5.5cm]{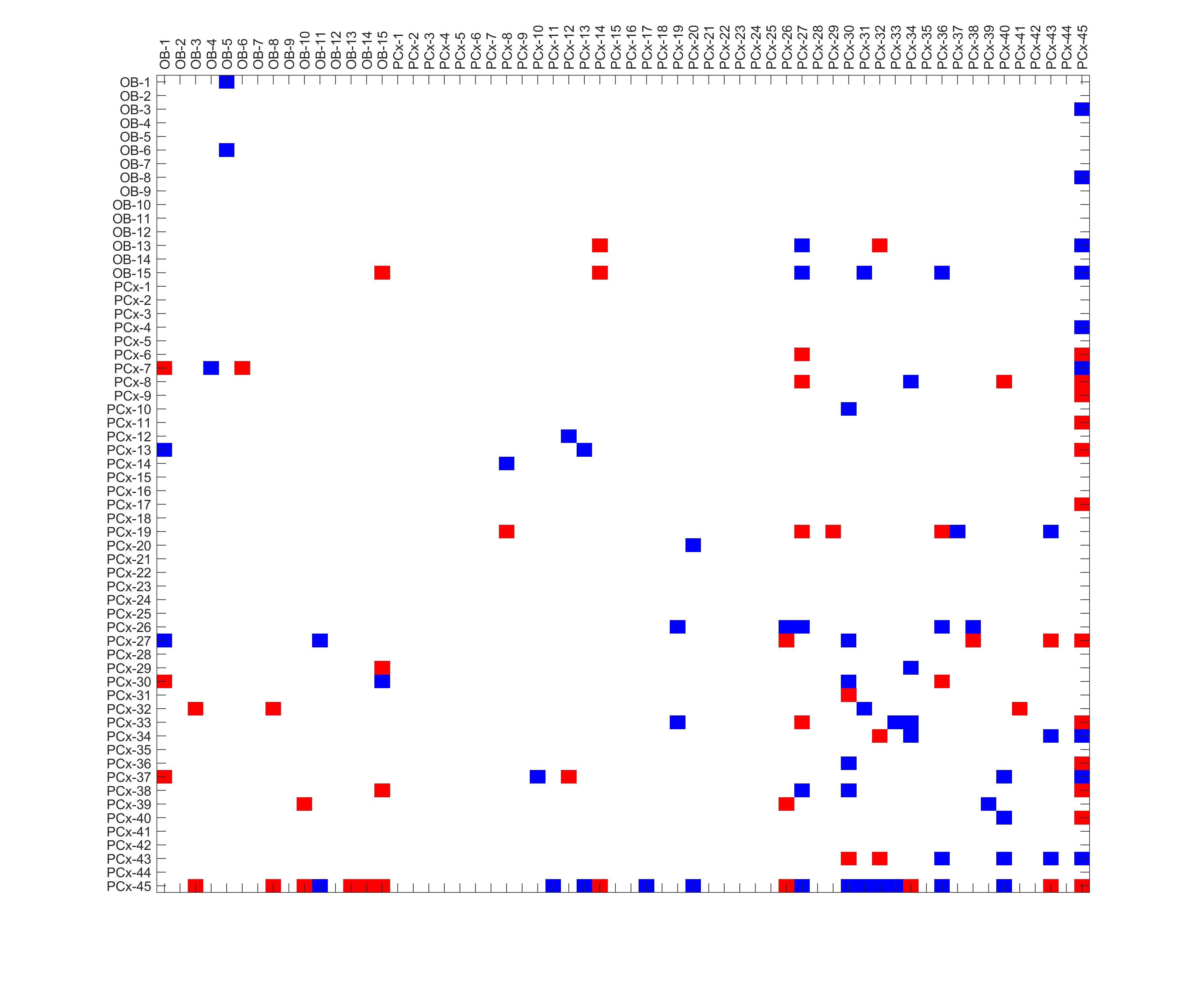}
\end{minipage}
\begin{minipage}[t]{0.30\textwidth}
\includegraphics[width = 5.5cm]{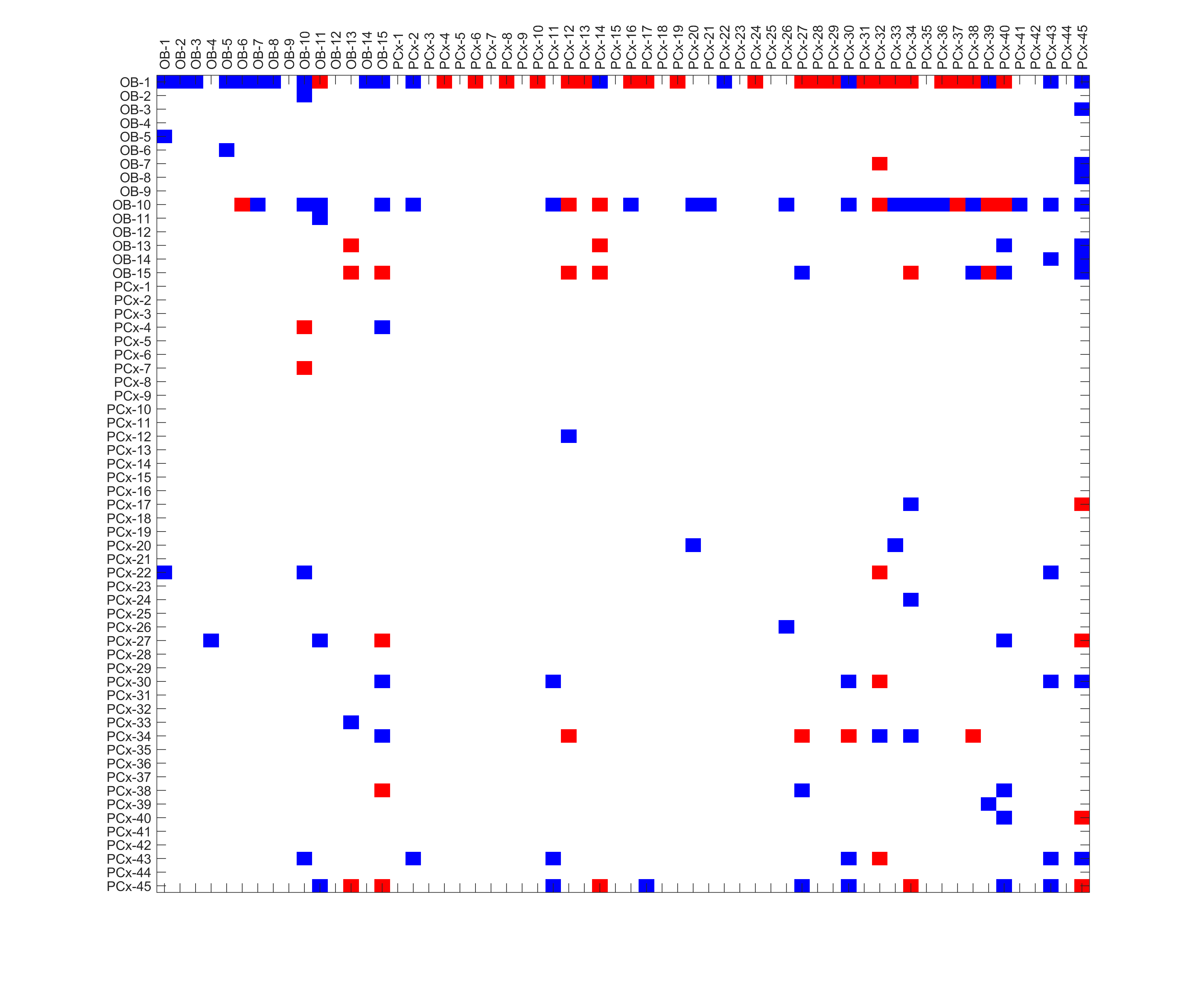}
\end{minipage}
\begin{minipage}[t]{0.30\textwidth}
\includegraphics[width = 5.5cm]{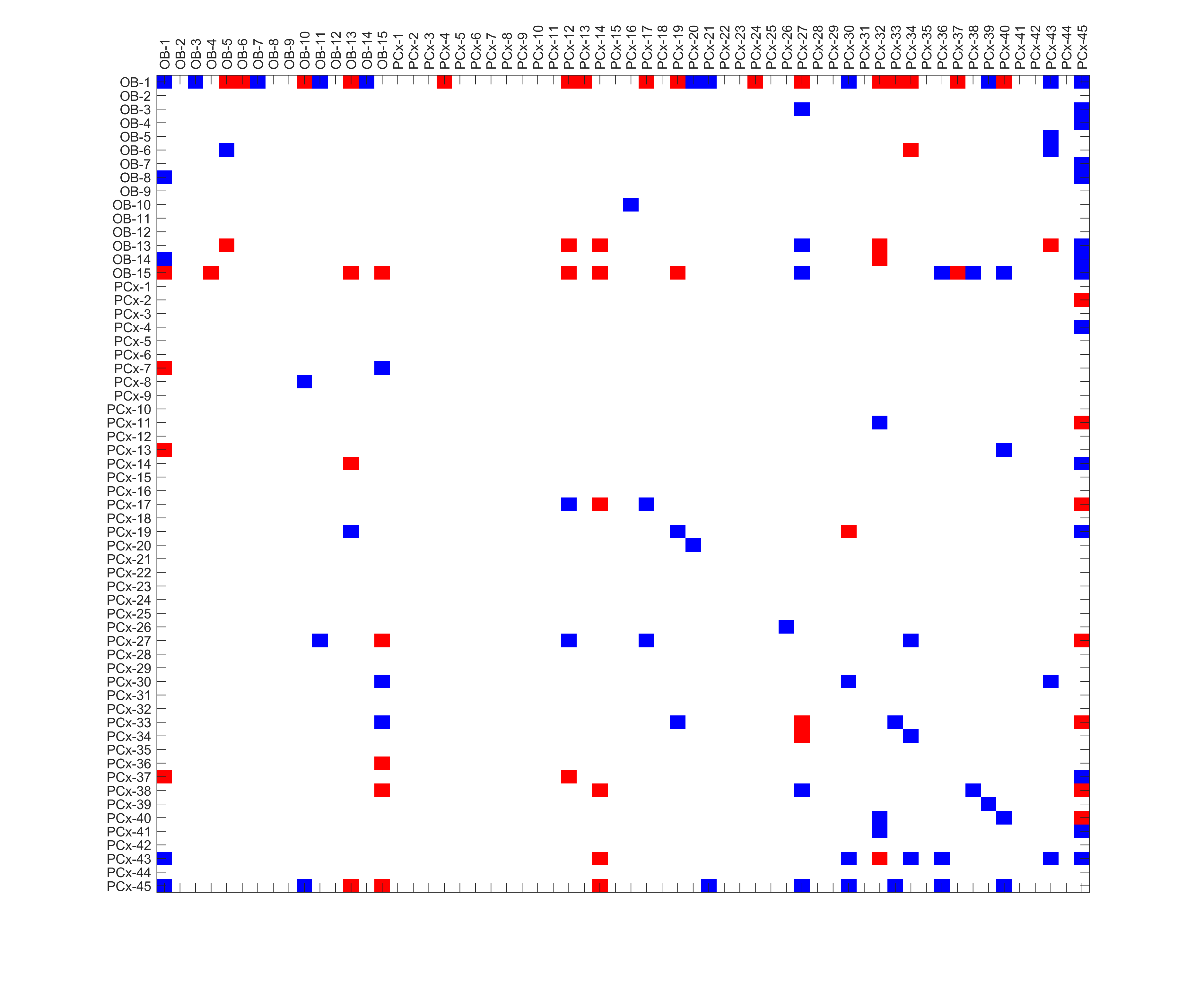}
\end{minipage}

\caption{Estimated Network among the OB cells and the PCx cells for laser intensity $0\ mW/mm^2$, $5\ mW/mm^2$ and $10\ mW/mm^2$ from left to right. OB cells are numbered 1-15 and PCx cells are numbered 16-60. A red pixel at $(i,j)$ indicates an inhibitory connection from cell $j$ to cell $i$ and a blue pixel at $(i,j)$ indicates an excitatory connection from cell $j$ to cell $i$.}
\label{FigureApplicationEstimatedConnection}
\end{figure}

Figure \ref{FigureApplicationEstimatedConnection} Shows the networks estimated using a point process graphical model for three groups. The bootstrapped graphical model estimated a sparser network with sparsity level $3.28\%$ for the $0\ mW/mm^2$ group, $4.08\%$ for the $5\ mW/mm^2$ group and $3.67\%$ for the $10\ mW/mm^2$ group. We observe a few apparent patterns. First, PCx cells are densely connected in all three groups but the $0mW/mm^2$ group has increased connectivity relative to other groups, suggesting that there may be a multi-layer excitatory-inhibitory neuron network in the piriform cortex to stabilize the odor signal.
Second, when the laser was applied to OB cells, the number of edges from other cells to the OB cells increase and the sign of if several connections changed, indicating that the laser affected the connectivity structure of OB cells and PCx cells and this effect was heterogenous. Finally, we observed that a few neurons fired less frequently and had little effect on other cells (such as OB-9, PCx-1, PCx-15, PCx-42, PCx-44, etc). Their role in the odor-processing mechanism needs to be further explored.

\section{Conclusion and Future Work}
In this paper, we have proposed the temporal point process graphical model, a new class of graphical models for learning the temporal dependencies among different components of a multivariate point process. We have shown that the locally independent structure of the process is fully encoded in the transfer matrix. We have provided a class of estimators with a non-asymptotic estimation error bound, which extends the classical estimators of temporal point process. Our model naturally includes the case of sparse temporal point process regression, where the predictors and the responses are not the same process. The performance of our estimator has been tested by simulations and a spike train data set.

Several challenges still remain in the analysis of the temporal point process graphical models. Using cross-validation for selecting tuning parameters is proved to be successful in the classical Lasso estimators (\citeauthor{chetverikov2021cross}, \citeyear{chetverikov2021cross}). However, a detailed theoretical analysis is needed to demonstrate its performance in our case. The criteria for optimal tuning parameter selection for estimation and variable selection are left for future research. The second challenge is to assess the uncertainty of the estimated parameters, which is less addressed in the existing work.

\bibliographystyle{plainnat}
\bibliographystyle{vancouver-authoryear} % Style BST file
\bibliography{TPPref}

\begin{thebibliography}{39}
\providecommand{\natexlab}[1]{#1}
\providecommand{\url}[1]{\texttt{#1}}
\expandafter\ifx\csname urlstyle\endcsname\relax
  \providecommand{\doi}[1]{doi: #1}\else
  \providecommand{\doi}{doi: \begingroup \urlstyle{rm}\Url}\fi

\bibitem[A{\"\i}t-Sahalia et~al.(2015)A{\"\i}t-Sahalia, Cacho-Diaz, and
  Laeven]{ait2015modeling}
Yacine A{\"\i}t-Sahalia, Julio Cacho-Diaz, and Roger~JA Laeven.
\newblock Modeling financial contagion using mutually exciting jump processes.
\newblock \emph{Journal of Financial Economics}, 117\penalty0 (3):\penalty0
  585--606, 2015.

\bibitem[Bacry et~al.(2013)Bacry, Delattre, Hoffmann, and Muzy]{bacry2013some}
Emmanuel Bacry, Sylvain Delattre, Marc Hoffmann, and Jean-Fran{\c{c}}ois Muzy.
\newblock Some limit theorems for hawkes processes and application to financial
  statistics.
\newblock \emph{Stochastic Processes and their Applications}, 123\penalty0
  (7):\penalty0 2475--2499, 2013.

\bibitem[Bacry et~al.(2015)Bacry, Mastromatteo, and Muzy]{bacry2015hawkes}
Emmanuel Bacry, Iacopo Mastromatteo, and Jean-Fran{\c{c}}ois Muzy.
\newblock Hawkes processes in finance.
\newblock \emph{Market Microstructure and Liquidity}, 1\penalty0 (01):\penalty0
  1550005, 2015.

\bibitem[Bacry et~al.(2020)Bacry, Bompaire, Ga{\"\i}ffas, and
  Muzy]{bacry2020sparse}
Emmanuel Bacry, Martin Bompaire, St{\'e}phane Ga{\"\i}ffas, and Jean-Francois
  Muzy.
\newblock Sparse and low-rank multivariate hawkes processes.
\newblock \emph{Journal of Machine Learning Research}, 21\penalty0
  (50):\penalty0 1--32, 2020.

\bibitem[Bolding and Franks(2018)]{2018Rec}
K.~A. Bolding and K.~M. Franks.
\newblock Recurrent cortical circuits implement concentration-invariant odor
  coding.
\newblock \emph{Science}, 361\penalty0 (6407):\penalty0 eaat6904--, 2018.

\bibitem[Br\'{e}maud and Massouli\'{e}(1996)]{Stability1996}
Pierre Br\'{e}maud and Laurent Massouli\'{e}.
\newblock Stability of nonlinear hawkes processes.
\newblock \emph{The Annals of Probability}, 24\penalty0 (3):\penalty0 1563 --
  1588, 1996.

\bibitem[Brémaud(1981)]{Pierre1981Point}
Pierre Brémaud.
\newblock \emph{Point Processes and Queues}.
\newblock Springer-Verlag, 1981.

\bibitem[Chen et~al.(2017{\natexlab{a}})Chen, Shojaie, Shea-Brown, and
  Witten]{2017chen}
Shizhe Chen, Ali Shojaie, Eric Shea-Brown, and Daniela Witten.
\newblock The multivariate hawkes process in high dimensions: Beyond mutual
  excitation.
\newblock \emph{arXiv preprint arXiv:1707.04928}, 2017{\natexlab{a}}.

\bibitem[Chen et~al.(2017{\natexlab{b}})Chen, Witten, and
  Shojaie]{chen2017nearly}
Shizhe Chen, Daniela Witten, and Ali Shojaie.
\newblock Nearly assumptionless screening for the mutually-exciting
  multivariate hawkes process.
\newblock \emph{Electronic journal of statistics}, 11\penalty0 (1):\penalty0
  1207, 2017{\natexlab{b}}.

\bibitem[Chetverikov et~al.(2021)Chetverikov, Liao, and
  Chernozhukov]{chetverikov2021cross}
Denis Chetverikov, Zhipeng Liao, and Victor Chernozhukov.
\newblock On cross-validated lasso in high dimensions.
\newblock \emph{The Annals of Statistics}, 49\penalty0 (3):\penalty0
  1300--1317, 2021.

\bibitem[Costa et~al.(2018)Costa, Graham, Marsalle, and Tran]{2018Costa}
Manon Costa, Carl Graham, Laurence Marsalle, and Viet~Chi Tran.
\newblock Renewal in hawkes processes with self-excitation and inhibition.
\newblock \emph{Advances in Applied Probability}, 52, 01 2018.

\bibitem[Daley and Vere-Jones(2007)]{daley2007introduction}
DJ~Daley and David Vere-Jones.
\newblock \emph{An Introduction to the Theory of Point Processes: Volume II:
  General Theory and Structure}.
\newblock Springer Science \& Business Media, 2007.

\bibitem[Didelez(2008)]{didelez2008graphical}
Vanessa Didelez.
\newblock Graphical models for marked point processes based on local
  independence.
\newblock \emph{Journal of the Royal Statistical Society: Series B (Statistical
  Methodology)}, 70\penalty0 (1):\penalty0 245--264, 2008.

\bibitem[Engelke and Hitz(2020)]{engelke2020graphical}
Sebastian Engelke and Adrien~S Hitz.
\newblock Graphical models for extremes.
\newblock \emph{Journal of the Royal Statistical Society: Series B (Statistical
  Methodology)}, 82\penalty0 (4):\penalty0 871--932, 2020.

\bibitem[Fox et~al.(2016)Fox, Short, Schoenberg, Coronges, and
  Bertozzi]{fox2016modeling}
Eric~W Fox, Martin~B Short, Frederic~P Schoenberg, Kathryn~D Coronges, and
  Andrea~L Bertozzi.
\newblock Modeling e-mail networks and inferring leadership using self-exciting
  point processes.
\newblock \emph{Journal of the American Statistical Association}, 111\penalty0
  (514):\penalty0 564--584, 2016.

\bibitem[Hansen et~al.(2015)Hansen, Reynaud-Bouret, and Rivoirard]{2015Hansen}
Niels Hansen, Patricia Reynaud-Bouret, and Vincent Rivoirard.
\newblock Lasso and probabilistic inequalities for multivariate point
  processes.
\newblock \emph{Bernoulli}, 21, 08 2015.

\bibitem[Hawkes(1971)]{1971Hawkes}
Alan~G Hawkes.
\newblock Spectra of some self-exciting and mutually exciting point processes.
\newblock \emph{Biometrika}, 58\penalty0 (1):\penalty0 83--90, 1971.

\bibitem[Hawkes and Oakes(1974)]{1974Hawkes}
Alan~G. Hawkes and David Oakes.
\newblock A cluster process representation of a self-exciting process.
\newblock \emph{Journal of Applied Probability}, 11\penalty0 (03):\penalty0
  493--503, 1974.

\bibitem[Kalisch and B{\"u}hlman(2007)]{DAGkalisch2007estimating}
Markus Kalisch and Peter B{\"u}hlman.
\newblock Estimating high-dimensional directed acyclic graphs with the
  pc-algorithm.
\newblock \emph{Journal of Machine Learning Research}, 8\penalty0 (3), 2007.

\bibitem[Liu(2013)]{Gaussianliu2013gaussian}
Weidong Liu.
\newblock Gaussian graphical model estimation with false discovery rate
  control.
\newblock \emph{The Annals of Statistics}, pages 2948--2978, 2013.

\bibitem[Maathuis et~al.(2018)Maathuis, Drton, Lauritzen, and
  Wainwright]{maathuis2018handbook}
Marloes Maathuis, Mathias Drton, Steffen Lauritzen, and Martin Wainwright.
\newblock \emph{Handbook of graphical models}.
\newblock CRC Press, 2018.

\bibitem[Negahban et~al.(2012)Negahban, Ravikumar, Wainwright, and
  Yu]{negahban2009unified}
Sahand~N Negahban, Pradeep Ravikumar, Martin~J Wainwright, and Bin Yu.
\newblock A unified framework for high-dimensional analysis of $ m $-estimators
  with decomposable regularizers.
\newblock \emph{Statistical science}, 27\penalty0 (4):\penalty0 538--557, 2012.

\bibitem[Ogata(1978)]{ogata1978asymptotic}
Yoshiko Ogata.
\newblock The asymptotic behaviour of maximum likelihood estimators for
  stationary point processes.
\newblock \emph{Annals of the Institute of Statistical Mathematics},
  30\penalty0 (2):\penalty0 243--261, 1978.

\bibitem[Ozaki(1979)]{ozaki1979maximum}
Tohru Ozaki.
\newblock Maximum likelihood estimation of {Hawkes'} self-exciting point
  processes.
\newblock \emph{Annals of the Institute of Statistical Mathematics},
  31\penalty0 (1):\penalty0 145--155, 1979.

\bibitem[Paninski et~al.(2007)Paninski, Pillow, and
  Lewi]{apppaninski2007statistical}
Liam Paninski, Jonathan Pillow, and Jeremy Lewi.
\newblock Statistical models for neural encoding, decoding, and optimal
  stimulus design.
\newblock \emph{Progress in brain research}, 165:\penalty0 493--507, 2007.

\bibitem[Perry and Wolfe(2013)]{perry2013point}
Patrick~O Perry and Patrick~J Wolfe.
\newblock Point process modelling for directed interaction networks.
\newblock \emph{Journal of the Royal Statistical Society: Series B (Statistical
  Methodology)}, 75\penalty0 (5):\penalty0 821--849, 2013.

\bibitem[Pillow et~al.(2008)Pillow, Shlens, Paninski, Sher, Litke,
  Chichilnisky, and Simoncelli]{apppillow2008spatio}
Jonathan~W Pillow, Jonathon Shlens, Liam Paninski, Alexander Sher, Alan~M
  Litke, EJ~Chichilnisky, and Eero~P Simoncelli.
\newblock Spatio-temporal correlations and visual signalling in a complete
  neuronal population.
\newblock \emph{Nature}, 454\penalty0 (7207):\penalty0 995--999, 2008.

\bibitem[Qiao et~al.(2019)Qiao, Guo, and James]{qiao2019functional}
Xinghao Qiao, Shaojun Guo, and Gareth~M James.
\newblock Functional graphical models.
\newblock \emph{Journal of the American Statistical Association}, 114\penalty0
  (525):\penalty0 211--222, 2019.

\bibitem[Shorack and Wellner(2009)]{shorack2009empirical}
Galen~R Shorack and Jon~A Wellner.
\newblock \emph{Empirical processes with applications to statistics}.
\newblock SIAM, 2009.

\bibitem[Tang and Li(2021)]{tang2021multivariate}
Xiwei Tang and Lexin Li.
\newblock Multivariate temporal point process regression.
\newblock \emph{Journal of the American Statistical Association}, 0\penalty0
  (0):\penalty0 1--16, 2021.

\bibitem[Tibshirani(1996)]{tibshirani1996regression}
Robert Tibshirani.
\newblock Regression shrinkage and selection via the lasso.
\newblock \emph{Journal of the Royal Statistical Society: Series B
  (Methodological)}, 58\penalty0 (1):\penalty0 267--288, 1996.

\bibitem[Van~de Geer et~al.(2013)Van~de Geer, B{\"u}hlmann,
  et~al.]{DAGvandegeer2013}
Sara Van~de Geer, Peter B{\"u}hlmann, et~al.
\newblock $\ell_0$-penalized maximum likelihood for sparse directed acyclic
  graphs.
\newblock \emph{Annals of Statistics}, 41\penalty0 (2):\penalty0 536--567,
  2013.

\bibitem[Vinci et~al.(2018)Vinci, Ventura, Smith, and Kass]{vinci2018adjusted}
Giuseppe Vinci, Val{\'e}rie Ventura, Matthew~A Smith, and Robert~E Kass.
\newblock Adjusted regularization in latent graphical models: Application to
  multiple-neuron spike count data.
\newblock \emph{The annals of applied statistics}, 12\penalty0 (2):\penalty0
  1068, 2018.

\bibitem[Yuan and Lin(2007)]{Gaussianyuan2007model}
Ming Yuan and Yi~Lin.
\newblock Model selection and estimation in the gaussian graphical model.
\newblock \emph{Biometrika}, 94\penalty0 (1):\penalty0 19--35, 2007.

\bibitem[Zhang et~al.(2016)Zhang, Chai, Guo, Gao, Devilbiss, and
  Zhang]{Chunming2016}
Chunming Zhang, Yi~Chai, Xiao Guo, Muhong Gao, David Devilbiss, and Zhengjun
  Zhang.
\newblock Statistical learning of neuronal functional connectivity.
\newblock \emph{Technometrics}, 58\penalty0 (3):\penalty0 350--359, 2016.

\bibitem[Zhao and Yu(2006)]{zhao2006model}
Peng Zhao and Bin Yu.
\newblock On model selection consistency of lasso.
\newblock \emph{The Journal of Machine Learning Research}, 7:\penalty0
  2541--2563, 2006.

\bibitem[Zheng et~al.(2018)Zheng, Aragam, Ravikumar, and
  Xing]{DAGzheng2018dags}
Xun Zheng, Bryon Aragam, Pradeep~K Ravikumar, and Eric~P Xing.
\newblock Dags with no tears: Continuous optimization for structure learning.
\newblock \emph{Advances in Neural Information Processing Systems}, 31, 2018.

\bibitem[Zhou et~al.(2013)Zhou, Zha, and Song]{zhou2013learning}
Ke~Zhou, Hongyuan Zha, and Le~Song.
\newblock Learning triggering kernels for multi-dimensional hawkes processes.
\newblock In \emph{International Conference on Machine Learning}, pages
  1301--1309. PMLR, 2013.

\bibitem[Zou(2006)]{zou2006adaptive}
Hui Zou.
\newblock The adaptive lasso and its oracle properties.
\newblock \emph{Journal of the American statistical association}, 101\penalty0
  (476):\penalty0 1418--1429, 2006.

\end{thebibliography}

\appendix

\section{Appendix}\label{app}
Here we show the proof of Theorem\ \ref{TheoremMainTheorem}. The proof of Theorem\ \ref{TheoremMainTheorem} is divided in three steps. First, we show that the loss function admits the restricted strong convexity condition in Theorem \ \ref{TheoremRSC}. Second, we show that the estimator has a convergence rate of $\sqrt{\frac{s\log (p\vee T)}{T}}$ as long as $\lambda_T$ dominate the statistical noise in Theorem\ \ref{TheoremLassoArgument}. In the last step, we show that $\lambda_T$ dominate the statistical noise with high probability in Lemma\ \ref{lemmalambda>}.

\begin{theorem}
\label{TheoremRSC}
Let \begin{equation}
\label{EquationEmpiricalRiskJ}
L_T(\mu_j,\beta_j) = \frac{1}{T}\int_0^T W_j(t)\Big(H_j(\mu_j^* + \langle\bbeta_j,\bx_j(t)\rangle)\mathrm{d} t - \langle\bbeta_j,\bx_j(t)\rangle\mathrm{d} N_j(t)\Big)
\end{equation}
denote the loss function. Under Assumption \ref{Assumption1} and \ref{Assumption2}, there exists constants $\kappa_1,\kappa_2 > 0$ such that for every $(\delta_\mu,\Delta^\tr) \in K(r,S)$,
\begin{align*}
& L_T(\mu^*_j + \delta_\mu,\bbeta^*_j+\Delta) - L_T(\mu^*_j ,\bbeta^*_j) - \langle\nabla_{\bbeta_j}L_T(\mu^*_j,\bbeta^*_j), \bDelta\rangle - \nabla_{\mu}L_T(\mu^*_j,\bbeta^*_j)\delta_\mu\\
\geq & \kappa_1\|(\delta_\mu,\bDelta^\tr)\|_2 \Big\{\|(\delta_\mu,\bDelta^\tr)\|_2 - \kappa_2(\tau+1)^2\varepsilon(T,p)\|(\delta_\mu,\bDelta^\tr)\|_1^2\Big\}
\end{align*}
holds with probability greater than $1 - \frac{c_1}{(p\vee T)^\tau}$, where $S$ is the support of $\bbeta^*_j$ and $K(r,S) = \{(\delta_\mu,\bDelta) \in \mbR^{p+1}, \|(\delta_\mu,\bDelta^\tr)\|_2 = r, \|(0,\bDelta_{S^c}^\tr)\|_1 \leq 4\|(\delta_\mu,\bDelta_S^\tr)\|_1\}$, $\varepsilon(T,p) =  \sqrt{\frac{\log(p\vee T)}{T}}.$
\end{theorem}

\begin{theorem}
\label{TheoremLassoArgument}
Under Assumption \ref{Assumption1} and \ref{Assumption2}, taking $\lambda_T = c_1\sqrt{\frac{\tau \log(p\vee T)}{T}}$ and conditioning on the event $\cup_{j=1}^p G_j$, where
\begin{align*}
G_j & = \Big\{\lambda_T \geq
 2 h_{\max}\Big|\frac{1}{T}\int_0^T W_j(t)\Big(h_j(\mu_j^* + \langle\bbeta,\bx_j(t)\rangle) \mathrm{d} t -\mathrm{d} N_j(t)\Big)\Big|\ \\
 &\text{and}\ \lambda_T \geq 2 h_{\max}\Big|\frac{1}{T}\int_0^T W_j(t)\Big(h_j(\mu_j^* + \langle\bbeta,\bx_j(t)\rangle) x_{j,k}(t) \mathrm{d} t - x_{j,k}(t)\mathrm{d} N_j(t)\Big) \Big|, \forall k \in [p]
  \Big\},
\end{align*}
we have
\begin{equation}
p^{-1/2}\|(\hat{\bmu},\hat{B}) - (\bmu^*,B^{*})\|_F \leq c_2 \sqrt{s}\lambda_T
\end{equation}
and
\begin{equation}
\|(\hat{\bmu},\hat{B}) - (\bmu^*,B^{*})\|_\infty \leq c_2 s\lambda_T
\end{equation}
hold with possibility greater than $1 - \frac{c_3}{(p\vee T)^{\tau}}$, where $s = 1 + \max_j \|\bbeta^*_j\|_0$.
\end{theorem}

\begin{lemma}
\label{lemmalambda>}
Under Assumption \ref{Assumption1} and \ref{Assumption2}, there exist positive constants $c_1,c_2 > 0$, with probability greater than $1 - \frac{c_1}{(p\vee T)^\tau}$ , we have
$$\Big|\frac{1}{T}\int_0^T W_j(t)\Big(h_j(\mu_j^* + \langle\bbeta,\bx_j(t)\rangle) x_{j,k}(t) \mathrm{d} t - x_{j,k}(t)\mathrm{d} N_j(t)\Big) \Big| \leq c_2 \sqrt{\frac{\tau\log(p\vee T)}{T}},\ \ \forall k \in [p]$$
and
$$\Big|\frac{1}{T} \int_0^T W_j(t)\Big(h_j(\mu_j^* + \langle\bbeta,\bx_j(t)\rangle) \mathrm{d} t -\mathrm{d} N_j(t)\Big)\Big|  \leq c_2 \sqrt{\frac{\tau\log(p\vee T)}{T}}.$$
\end{lemma}

\subsection{Proof of Theorem \ref{TheoremRSC} }
We will show the restricted strong convexity for the loss function.

\textit{proof.} With a little abuse of notation, we let $\bx_j(t)$ denote the centralized influence function $\bx_j(t) - \mbE[\bx_j(t)]$ and $\mu^*_j$ denote $\mu^*_j + \sum_{k \in S} \beta_{j,k}^*\mbE[x_{j,k}(t)]$.

For $\bDelta \in \mbR^p$, define $\delta L_T(\delta_{\mu},\mu^*_j,\bDelta,\bbeta^*_j) = L_T(\mu^*_j+\delta_{\mu},\bbeta_j^*+\bDelta) - L_T(\mu_j^*,\bbeta_j^*) - \langle\nabla_{\bbeta} L_T(\mu_j^*,\bbeta_j^*),\bDelta\rangle - \delta_{\mu}\frac{\partial}{\partial \mu}L_T(\mu,\bbeta_j^*)\Big|_{\mu^*_j}$. By the Taylor expansion, there exists some $v \in [0,1]$ such that
$$\delta L_T(\delta_{\mu},\mu^*_j,\bDelta,\bbeta^*_j) = \frac{1}{T}\int_0^T W_j(t) h_j'\Big(\mu^*_j + \langle\bbeta_j^*,\bx_j(t)\rangle + v\big(\delta_\mu+\langle\bDelta,\bx_j(t)\rangle\big)\Big)(\delta_\mu + \langle\bDelta,\bx_j(t)\rangle)^2 \d t.$$

Let $\gamma$ be positive numbers to be determined later, $0 <\gamma \leq R$, where $R$ is the constant in Assumption \ref{Assumption2}. Define the functions $\phi_{\gamma}(u) = u^2\mbI(|u|\leq \gamma/2) + (\gamma-u)^2 \mbI(\gamma/2 \leq |u|\leq \gamma)$ and $\zeta_\gamma(u) = u\mbI(|u|\leq \gamma)$.

Note that
\begin{align*}
(\mu_j^*+\langle\bbeta^*_j,\bx_{j}(t)\rangle)\mbI(|\mu_j^*+\langle\bbeta_j^*,\bx_{j}(t)\rangle| \leq R) = \zeta_R(\mu_j^*+\langle\bbeta_j^*,\bx_{j}(t)\rangle),\\
(\delta_{\mu}+\langle\bDelta,\bx_{j}(t)\rangle)\mbI(|\delta_{\mu}+\langle\bDelta,\bx_{j}(t)\rangle|\leq \gamma) = \zeta_\gamma(\delta_{\mu}+\langle\bDelta,\bx_{j}(t)\rangle),\\
(\delta_{\mu}+\langle\bDelta,\bx_{j}(t)\rangle)^2 \mbI(|\delta_{\mu}+\langle\bDelta,\bx_{j}(t)\rangle|\leq \gamma) \geq \phi_\gamma(\delta_{\mu}+\langle\bDelta,\bx_{j}(t)\rangle).
\end{align*}
Then
\begin{align*}
 \delta L_T(\delta_{\mu},\mu^*_j,\bDelta,\bbeta^*_j)
\geq & \frac{1}{T}\int_0^T W_j(t)h_j'\Big(\mu^*_j + \langle\bbeta_j^*,\bx_j(t)\rangle + v\big(\delta_\mu+\langle\bDelta,\bx_j(t)\rangle\big)\Big)\cdot\mbI(|\mu^*_j + \langle\bbeta^*_j,\bx_j(t)\rangle| \leq R)\\
\cdot&  (\delta_\mu + \langle\bDelta,\bx_j(t)\rangle)^2\cdot\mbI(|\delta_\mu+\langle\bDelta,\bx_j(t)\rangle|\leq \gamma)\d t\\
\geq  & \frac{1}{T}\int_0^T W_j(t)h_j'\Big(\mu^*_j + \langle\bbeta_j^*,\bx_j(t)\rangle + v\big(\delta_\mu+\langle\bDelta,\bx_j(t)\rangle\big)\Big)\cdot\mbI(|\mu^*_j + \langle\bbeta^*_j,\bx_j(t)\rangle| \leq R)\\
\cdot&  \phi_\gamma(\delta_\mu + \langle\bDelta,\bx_j(t)\rangle)\d t\\
\geq & C_{h_j}(R)\frac{1}{T}\int_0^T\phi_{\gamma}(\delta_{\mu}+\langle\bDelta,\bx_{j}(t)\rangle)\cdot\mbI(|\mu_j^*+\langle\bbeta_j^*,\bx_{j}(t)\rangle| \leq R) \d t,
\end{align*}
where $C_{h_j}(R) = \sigma_3\inf\limits_{|x|\leq 2R} {h_j}'(x)$ is a constant by Assumption \ref{Assumption2}. In this vein, it suffices to bound
$$ \frac{1}{T}\int_0^T\phi_{\gamma}(\delta_{\mu}+\langle\bDelta,\bx_{j}(t)\rangle)\cdot\mbI(|\mu_j^*+\langle\bbeta_j^*,\bx_{j}(t)\rangle| \leq R) \d t.$$

Let $\gamma(r) = Rr$, Note that $\phi_c(cz) = c^2\phi_1(z)$ for any $c > 0$ and $z \in \mbR$. Then for $\|(\delta_{\mu},\bDelta^\tr)\|_2 = r$, $\phi_{\gamma(r)}(\delta_{\mu}+\langle\bDelta,\bx_{j,m}\rangle) = r^2\phi_R(\delta_{\mu}/r + \langle\bDelta/r,\bx_{j,m}\rangle)$, and

\begin{align*}
& \frac{1}{T}\int_0^T \phi_{\gamma(r)} (\delta_{\mu}+\langle\bDelta,\bx_{j}(t)\rangle) \cdot\mbI(|\mu_j^*+\langle\bbeta_j^*,\bx_{j}(t)\rangle|\leq R) \d t \\
= & \frac{r^2}{T}\int_0^T \phi_R(\delta_{\mu}/r+\langle\bDelta/r,\bx_{j}(t)\rangle)\cdot\mbI(|\mu_j^*+\langle\bbeta_j^*,\bx_{j}(t)\rangle| \leq R)\d t,
\end{align*}
which implies that it suffices to show, there exist strictly positive constants $\kappa_1$ and $\kappa_2$ which depends only on the $\Sigma_j$ and $h_j(\cdot)$ such that
$$\frac{1}{T}\int_0^T \phi_R(\delta_{\mu}/r + \langle\bDelta/r,\bx_{j}(t)\rangle)\cdot \mbI(|\mu_j^*+\langle\bbeta_j^*,\bx_{j}(t)\rangle|\leq R)\d t \geq \kappa_1 \Big\{1 - \kappa_2(\tau+1)^2\varepsilon(T,p) \cdot \frac{\|(\delta_{\mu},\bDelta^\tr)\|_1^2}{\|(\delta_{\mu},\bDelta^\tr)\|_2}\Big\}.$$
From this perspective, we only need to prove the inequality when $\|(\delta_{\mu},\bDelta^\tr)\|_2 = 1$.

Denote by $\mbS_2(1)$ the unit sphere and let $\mbS_1(\varpi) = \{(\delta_{\mu},\bDelta^\tr) \in \mbR^{p+1} : \|(\delta_{\mu},\bDelta^\tr)\|_1 = \varpi\}$. Let $f_{(\delta_{\mu},\bDelta^\tr)}(\bx_{j}) = \frac{1}{T}\int_0^T\phi_R(\delta_{\mu} + \langle\bDelta,\bx_{j}(t)\rangle) \cdot\mbI(|\mu_j^*+\langle\bbeta_j^*,\bx_{j}(t)\rangle| \leq R) \d t$. For each $\varpi > 0$,
it suffices to show that
$$\mcE_{\varpi} = \bigg\{f_{(\delta_{\mu},\bDelta^\tr)}(\bx_{j}) < \kappa_1\bigg(1 - \kappa_2(\tau+1)^2\varepsilon(T,p)\varpi^2\bigg),\ \text{for some}\ (\delta_{\mu},\bDelta^\tr) \in \mbS_1(\varpi)\cap\mbS_2(1)\bigg\}$$
occurs with small probability. Note that $\mbS_1(\varpi)\cap\mbS_2(1) \neq \emptyset$ only if $1 \leq \varpi \leq {\sqrt{p+1}}$.

Recall our Assumption \ref{Assumption2} (2), we have
$$\mbE[f_{(\delta_{\mu},\bDelta^\tr)}(\bx_{j})] > \sigma_1 > 0$$
holds for some $\sigma_1$.

By Lemma \ref{ConcentrationOfQuadratic}, we have for any $\varpi^2 \in [p]$,
$$\mbP\Big(\sup_{(\delta_{\mu},\bDelta^\tr) \in \mbS_2(1)\cap \mbS_1(\varpi)} \Big|\frac{1}{T}\int_0^T |\delta_\mu + \langle\bDelta,\bx_j(t)\rangle|^2 - \mbE|\delta_\mu+\langle\bDelta,\bx_j\rangle|^2\Big| \geq c_4 \tau^2 \varpi^2\epsilon(T,p)\Big) \leq \frac{c_5}{(p\vee T)^\tau},$$
where
$$ \varepsilon(T,p) = \sqrt{\frac{\log(p\vee T)}{T}}.$$
Since $\phi_R(\cdot)$ is $R$-Lipschitz function, we have

$$\mbP\Big(\sup_{(\delta_{\mu},\bDelta^\tr) \in \mbS_2(1)\cap \mbS_1(\varpi)}\Big| f_{(\delta_{\mu},\bDelta^\tr)}(\bx_{j}(t)) - \mbE[f_{(\delta_{\mu},\bDelta^\tr)}(\bx_{j})]\Big| \geq  c_4\tau^2R\varpi^2  \varepsilon(T,p)  \Big) \leq \frac{c_5}{(p\vee T)^\tau}.$$

Then
\begin{align*}
f_{(\delta_{\mu},\bDelta^\tr)}(\bx_{j}) & \geq \mbE[f_{(\delta_{\mu},\bDelta^\tr)}(\bx_{j})] - \sup_{(\delta_{\mu},\bDelta^\tr) \in \mbS_2(1)\cap\mbS_1(\varpi)}\Big| f_{(\delta_{\mu},\bDelta^\tr)}(\bx_{j}) - \mbE[f_{(\delta_{\mu},\bDelta^\tr)}(\bx_{j})]\Big|\\
& \geq \sigma_1 -  c_4\tau^2R\varpi^2  \varepsilon(T,p) \\
& = \sigma_1 \Big(1 - \frac{c_4\tau^2R\varpi^2}{\sigma_1}\varepsilon(T,p)\Big)
\end{align*}
holds with probability greater than $1 - \frac{c_5}{(p\vee T)^\tau}$.

The last step is "peeling". For any $\varpi^2 \in \{1,2,\cdots, p+1\}$, we have
$$\mbP\Big\{ f_{(\delta_{\mu},\bDelta^\tr)}(\bx_{j}) < \sigma_1 \Big(1 - \frac{c_4\tau^2R\varpi^2}{\sigma_1}\varepsilon(T,p)\Big),\ \text{for some}\ (\delta_{\mu},\bDelta^\tr) \in \mbS_1(\varpi)\cap \mbS_2(1)\Big\} \leq \frac{c_5}{(p\vee T)^\tau}.$$
Then we have
\begin{align*}
& \mbP\bigg\{ f_{(\delta_{\mu},\bDelta^\tr)}(\bx_{j}) < \frac{\sigma_1}{2} \Big(1 - \frac{c_4\tau^2R}{\sigma_1}\varepsilon(T,p)\Big),\ \text{for some}\ (\delta_{\mu},\bDelta^\tr) \in \mbS_2(1)\bigg\}\\
\leq & \sum_{i=1}^{p+1}\mbP\bigg\{ f_{(\delta_{\mu},\bDelta^\tr)}(\bx_{j}) < \sigma_1 \Big(1 - \frac{c_4\tau^2R}{\sigma_1}\varepsilon(T,p)\cdot i\Big),\ \text{for some}\ (\delta_{\mu},\bDelta^\tr) \in \mbS_1(\sqrt{i})\cap\mbS_2(1)\bigg\}\\
\leq & \frac{c_6}{(p\vee T)^{\tau-1}}.
\end{align*}

Then we conclude that by choosing $\kappa_1 = \frac{\sigma_1}{2}$ and $\kappa_2 = \frac{c_4 R}{\sigma_1}$ with some constant $R$, the following
\begin{align*}
\delta L_T(\delta_{\mu},\mu^*_j,\bDelta,\bbeta^*_j) & \geq \kappa_1\|(\delta_{\mu},\bDelta^\tr)\|_2\Big\{\|(\delta_{\mu},\bDelta^\tr)\|_2 - \kappa_2(\tau+1)^2\varepsilon(T,p)\|(\delta_{\mu},\bDelta^\tr)\|_1^2\Big\}\\
\end{align*}
holds uniformly for all $\|(\delta_{\mu},\bDelta^\tr)\|_2 \leq 1$ with probability greater than $1 - \frac{c_6}{(p\vee T)^\tau}$. \hfill$\Box$
\subsection{Proof of Theorem \ref{TheoremLassoArgument} }

Let $\mcF(\delta_{\mu},\bDelta) = L_T(\mu_j^*+\delta_\mu,\bbeta_j^* + \bDelta) -  L_T(\mu_j^*,\bbeta_j^*) + \lambda_T(\|(\mu_j^*+\delta_{\mu},(\bbeta_j^*+\bDelta)^\tr)\|_1 - \|(\mu_j^*,\bbeta_j^{*\tr})\|_1)$ be the difference of penalized loss function. Let $(\hat{\mu}_j,\hat{\bbeta}_j)$ be the optimal solution to the penalized loss function minimization and define $\hat{\delta}_{\mu} = \hat{\mu}_j - \mu_j^*, \hat{\bDelta} = \hat{\bbeta}_j - \bbeta_j^*$. We first show some properties of the global optimizer $(\hat{\mu}_j,\hat{\bbeta}_j)$.

Notice that $L_T(\cdot,\cdot)$ is a convex function. Then
$$L_T(\mu_j^*+\hat{\delta}_{\mu},\bbeta_j^*+\hat{\bDelta}) - L_T(\mu_j^*,\bbeta_j^*) \geq \langle\nabla_{\bbeta}L_T(\mu_j^*,\bbeta_j^*),\hat{\bDelta}\rangle + \nabla_{\mu}L_T(\mu_j^*,\bbeta_j^*)\cdot\hat{\delta}_{\mu} \geq -\frac{\lambda_T}{2}\|(\hat{\delta}_{\mu},\hat{\bDelta}^{\tr})\|_1.$$
$\mcF(\hat{\delta}_{\mu},\hat{\bDelta}) \leq 0$ gives that
$$-\frac{\lambda_T}{2}\|\hat{\bDelta}\|_1 \leq \lambda_T\Big(\|(\mu_j^*,\bbeta_j^{*\tr})\|_1 - \|(\mu_j^*+\hat{\delta}_{\mu}, (\bbeta_j^*+\hat{\bDelta})^\tr)\|_1\Big).$$
Owing to
\begin{align*}
\|(\mu_j^*,\bbeta_j^{*\tr})\|_1 & = \|(\mu_j^*,\bbeta^{*\tr}_{j,S})\|_1 \\
\|(\hat{\delta}_{\mu},\hat{\bDelta}^\tr)\|_1 & = \|(\hat{\delta}_{\mu},\hat{\bDelta}^\tr_S)\|_1 + \|(0,\hat{\bDelta}^\tr_{S^C})\|_1 \\
\|(\mu_j^*+\hat{\delta}_\mu, (\bbeta_j^*+\hat{\bDelta})^\tr)\|_1 & = \|(\mu_j^*+\hat{\delta}_\mu, (\bbeta_{j,S}^*+\hat{\bDelta}_S)^\tr)\|_1 + \|(0, (\bbeta_{j,S^C}^*+\hat{\bDelta}_{S^C})^\tr)\|_1\\
\|(\hat{\delta}_\mu,\hat{\bDelta}_S^{\tr})\|_1 & \geq \|(\mu_j^*,\bbeta^{*\tr}_{j,S})\|_1 - \|(\mu_j^*+\hat{\delta}_\mu,(\bbeta_j^* + \hat{\bDelta})^\tr_S)\|_1,
\end{align*}
where the subscript $S$ denotes that the vector is restricted to the set $S$, we have
$$\|\hat{\bDelta}_{S^C}\|_1 \leq 3 \|(\hat{\delta}_\mu,\hat{\bDelta}_S^\tr)\|_1\ \text{and}\ \|(\hat{\delta}_{\mu},\hat{\bDelta}^\tr)\|_1 \leq 4 \|(\hat{\delta}_{\mu},\hat{\bDelta}_S^\tr)\|_1 \leq 4\sqrt{s}\|(\hat{\delta}_{\mu},\hat{\bDelta}_S^\tr)\|_2.$$

Consider the set $K(r,S) = \{(\delta_{\mu},\bDelta^\tr) \in \mbR^{p+1}, \|(\delta_{\mu},\bDelta^\tr)\|_2 = r, \|(0,\bDelta^\tr_{S^C})\|_1 \leq 4 \|({\delta}_{\mu},{\bDelta}_S^\tr)\|_1\}$ for $0 < r < 1$. With the restricted strong convexity condition provided by Theorem \ref{TheoremRSC}, we have for any $(\delta_\mu,\bDelta^\tr) \in K(r,S)$,
\begin{align*}
&L_T(\mu_j^*+\delta_{\mu}, \bbeta_j^*+\bDelta) - L_T(\mu_j^*,\bbeta_j^*) - \langle\nabla_{\bbeta} L_T(\mu_j^*,\bbeta_j^*),\bDelta\rangle - \nabla_{\mu} L_T(\mu_j^*,\bbeta_j^*)\delta_{\mu}\\
\geq & \kappa_1 \|(\delta_{\mu},\bDelta^\tr)\|_2 \Big\{\|(\delta_{\mu},\bDelta^\tr)\|_2 - \kappa_2\tau^2\varepsilon(T,p)\|(\delta_{\mu},\bDelta^\tr)\|_1^2\Big\}\\
\geq & \kappa_1 \|(\delta_{\mu},\bDelta^\tr)\|_2 \Big\{\|(\delta_{\mu},\bDelta^\tr)\|_2 - s\kappa_2\tau^2\varepsilon(T,p)\|(\delta_{\mu},\bDelta^\tr)\|_2^2\Big\}
\end{align*}
holds with probability greater than $1 - \frac{c_6}{(p\vee T)^\tau}$, where $\epsilon(T,p) = \sqrt{\frac{\log (p \vee T)}{T}}$.
Further,\ condition on the event \begin{align*}
G_j & = \Big\{\lambda_T \geq
 2 h_{\max}\Big|\frac{1}{T}\int_0^T W_j(t)\Big(h_j(\mu_j^* + \langle\bbeta,\bx_j(t)\rangle) \mathrm{d} t -\mathrm{d} N_j(t)\Big)\Big|\ \\
 &\text{and}\ \lambda_T \geq 2 h_{\max}\Big|\frac{1}{T}\int_0^T W_j(t)\Big(h_j(\mu_j^* + \langle\bbeta,\bx_j(t)\rangle) x_{j,k}(t) \mathrm{d} t - x_{j,k}(t)\mathrm{d} N_j(t)\Big) \Big|, \forall k \in [p]
  \Big\},
\end{align*} note that $\|(\hat{\delta}_\mu,\hat{\bDelta}^\tr)\|_2 = r < 1$, we can derive that

\begin{align*}
\mcF(\hat{\delta}_{\mu},\hat{\bDelta}) & = L_T(\mu_j^*+\hat{\delta}_{\mu}, \bbeta^*+\hat{\bDelta}) - L_T(\mu_j^*,\bbeta_j^*)+ \lambda_T\Big(\|(\mu_j^*+\hat{\delta}_\mu,(\bbeta_j^*+\hat{\bDelta})_S^\tr)\|_1 + \|(0,\hat{\bDelta}^\tr_{S^C})\|_1 - \|(\mu_j^*,\bbeta_j^{*\tr})\|_1\Big)\\
& \geq -\frac{\lambda_T}{2}\|(\hat{\delta}_\mu,\hat{\bDelta}^\tr)\|_1 + \kappa_1 \|(\hat{\delta}_\mu,\hat{\bDelta}^\tr)\|_2 \Big\{\|(\hat{\delta}_\mu,\hat{\bDelta}^\tr)\|_2 -s\kappa_2 \tau^2\varepsilon(T,p)\|(\hat{\delta}_\mu,\hat{\bDelta}^\tr)\|_2^2\Big\}\\
&+ \lambda_T(\|(0,\hat{\bDelta}_{S^C}^\tr)\|_1 - \|(\hat{\delta}_\mu,\hat{\bDelta}_S^\tr)\|_1) \\
& \geq \kappa_1 \|(\hat{\delta}_\mu,\hat{\bDelta}^\tr)\|_2 \Big\{\|(\hat{\delta}_\mu,\hat{\bDelta}^\tr)\|_2 -s\kappa_2 \tau^2\varepsilon(T,p)\|(\hat{\delta}_\mu,\hat{\bDelta}^\tr)\|_2^2\Big\} - \frac{3\sqrt{s} \lambda_T}{2}\|(\hat{\delta}_\mu,\hat{\bDelta}^\tr)\|_2\\
& \geq \Big(\frac{\kappa_1}{2}\|(\hat{\delta}_\mu,\hat{\bDelta}^\tr)\|_2 - 3\sqrt{s}\lambda_T\Big)\|(\hat{\delta}_\mu,\hat{\bDelta}^\tr)\|_2,
\end{align*}
where the last inequality comes form $(\hat{\delta}_\mu,\hat{\bDelta}^\tr) \in K(r,S), s\kappa_2 \tau^2\varepsilon(T,p) < \frac{1}{2}$ for $T$ large enough and the choice of $\lambda_T$. To derive the final result, we need to show that $\|(\hat{\delta}_\mu,\hat{\bDelta}^\tr)\|_2 = r < 1$. Consider the choice of $r = \frac{6\sqrt{s}\lambda_T}{\kappa_1} < 1$. In this case, $\mcF(\delta_\mu,\bDelta) > 0$ for every $(\delta_{\mu},\bDelta^\tr) \in K(r,S)$. If $\|(\hat{\delta}_\mu,\hat{\bDelta}^\tr)\|_2 > r$, then there exists a vector $(a\hat{\delta}_\mu,a\hat{\bDelta}^\tr)$ with $0 < a < 1$ and $\|(a\hat{\delta}_\mu,a\hat{\bDelta}^\tr)\|_2 = \delta$, As $\mcF$ is convex, then
$$\mcF(a\hat{\delta}_\mu,a\hat{\bDelta}) \leq a \mcF(\hat{\delta}_\mu,\hat{\bDelta}) + (1-a)\mcF(0) = a\mcF(\hat{\delta}_\mu,\hat{\bDelta}) \leq 0.$$
Note that $(a\hat{\delta}_\mu,a\hat{\bDelta}^\tr) \in K(r,S)$. We come to a contradictory and we conclude that $\|\hat{\delta}_\mu,\hat{\bDelta}^\tr\|_2 \leq \frac{6\sqrt{s}\lambda_T}{\kappa_1}$.
\hfill$\Box$
\subsection{Proof of Lemma \ref{lemmalambda>}}

Recall the well known Bernstein type inequalities for point process which can be found in \citeauthor{shorack2009empirical}(\citeyear{shorack2009empirical}):
\begin{equation}
\label{EquationBernstein}\mbP\Big(|M_{T}| \geq \sqrt{2vx} + \frac{Bx}{3}\ \text{and}\ \int_{0}^T g^2(t)\d \Lambda(t)\leq v \ \text{and}\ \sup_{t\in[0,T]}|g(t)| \leq B\Big) \leq 2e^{-x},\end{equation}
where $g(t)$ is a predictable process, $\Lambda(t)$ is the compensator of the point process $N(t)$ and $M_T = \int_0^T g(t)(\d N(t) - \d \Lambda(t))$.

Conditioning on $C_\tau$ defined in Lemma \ref{lemmaSetC} and define the predictable process $g_{j,k}(t) = W_j(t)x_{j,k}(t)$, $\Lambda(t) = \Lambda_j(t) = \int_0^t h_j(\mu_j + \langle \bbeta_j,\bx_j(s)\rangle)\d s,\ M_{k,T} = \int_0^T g_{j,k}(t)(\d N_j(t) - \d \Lambda_j(t))$ and $g_{j,k}(t)$ is a predictable process with $\int_0^T g_{j,k}^2(t) \d \Lambda_j(t) \leq \sigma^2_4\int_0^t x^2_{j,k}(s) \d \Lambda_j(t)$. With Lemma \ref{Wangxx^Tconcentration}, we have
\begin{align*}
\int_0^T g_{j,k}^2(t) \d \Lambda_j(t)
&\leq h_{\max}\sigma^2_4\cdot\Big(  \Big|\int_0^T x^2_{j,k}(t)\d t - T\mbE[x_{j,k}^2(t)]\Big| + T\mbE[x_{j,k}^2(t)] \Big)\\
& \leq c_2T(\mbE[x^2_{j,k}(t)] + 1)
\end{align*}
holds simultaneously for all $k \in [p]$ with probability greater than $1 - \frac{c_1}{(p \vee T)^\tau}$. On the event $C_\tau$ in Lemma \ref{lemmaSetC}, we have $\sup_{k,t\in[0,T]}|g_{j,k}(t)| = \sup_{k,t\in[0,T]}|x_{j,k}(t)| \leq (\tau+2)\kappa_{\max}\log(p\vee T)$.

Notice that
$$|M_{k,T}| = |\int_0^T g_{j,k}(t)\big(\d N_j(t) - \d \Lambda_j(t)\big)| .$$
Let $v = c_2T(\mbE[x^2_{j,k}(t)] + 1)$ and $B = (\tau+2)\kappa_{\max}\log(p\vee T)$ in \eqref{EquationBernstein}, we have
$$\mbP\Big(|M_{k,T}| \geq \sqrt{c_2T(\mbE[x^2_{j,k}(t)] + 1) x }+\frac{(\tau+2)\kappa_{\max}\log(p\vee T)}{3}x\Big) \leq 2e^{-x}.$$
and
$$\mbP\Big(|M_{k,T}| \geq \sqrt{c_2T(\mbE[x^2_{j,k}(t)] + 1) x }+\frac{(\tau+2)\kappa_{\max}\log(p\vee T)}{3}x,\ \forall k \in [p]\Big) \leq 2pe^{-x}.$$

Take $x= (\tau+2)\log (p \vee T)$ and notice that $\sqrt{c_2T(\mbE[x^2_{j,k}(t)] + 1) x}$ dominate $\frac{(\tau+2)\kappa_{\max}\log(p\vee T)}{3}x$, we will have

$$\Big|\frac{1}{T}\int_0^T h_j(\mu_j^* + \langle\bbeta,\bx_j(t)\rangle) x_{j,k}(t) \d t - x_{j,k}(t)\d N_j(t) \Big| \leq c_2 \sqrt{\frac{\tau\log(p\vee T)}{T}},\ \ \forall k \in [p]$$
holds for all $k \in [p]$ with probability greater than $1 - \frac{c_2}{(p \vee T)^\tau}$ where $c_1,c_2$ are constants do not depend on $p$ or $T$. The second inequation can be proved similarly. \hfill$\Box$

\subsection{Ancillary Results}
\begin{lemma}
\label{lemmaSetB}
$N_j(T)$ is the number of event on the node $j$ in the time interval $[0,T]$ and there exists a constant $C$ that only depends on $h_{\max}$ that with probability greater than $1 - o(e^{-C_0T})$, we have
$$N_j(T) \leq C e h_{\max}T,\ \forall j = 1,\cdots,p.$$

\end{lemma}
\textit{proof:}
By the idea of Poisson thinning, we have
$$ N_j(T) \leq \tilde{N}_j(T)$$
and for $C \geq 2$,
\begin{align*}
\mbP(\tilde{N}_j(T) > Ceh_{\max}T,\ \forall j \in [p]) & \leq p\Big(\sum_{k = [Ceh_{\max}T]+1}^{\infty} \frac{(Th_{\max})^k}{k!}e^{-Th_{\max}}\Big) \\
& \leq \frac{2p(Th_{\max})^{CeT h_{\max}}}{([Ceh_{\max}T]+1)!}e^{-Th_{\max}}\\
& \lesssim \frac{p(Th_{\max})^{CeT h_{\max}}}{(CeTh_{\max})^{CeTh_{\max}}e^{-CeTh_{\max}}\sqrt{CT}}e^{-Th_{\max}} \\
& = \frac{1}{\sqrt{CT}}\exp\Big\{ - (Ce \log C - C + 1)T h_{\max} + \log p\Big\} .
\end{align*}
Because $\log(p) = o(T)$, we can pick $C$ only depending on $h_{\max}$ such that with probability greater than $1 - o(e^{-C T})$, we have
$$N_j(T) \leq C e h_{\max}T,\ \forall j = 1,\cdots,p.$$
\hfill $\Box$

Define the event $B = \{N_j(T) \leq Ceh_{\max} T,\ \forall j = 1,\cdots, p\}$ and $\mbP(B) = 1 - o(e^{-CT})$.
\begin{lemma}
\label{lemmaSetC}
Under Assumption \ref{Assumption1}, we have
\begin{equation}
\label{boundforx}
\mbP\Big(\sup_{k \in [p], t \in [0,T]}|x_{j,k}(t)| > \kappa_{\max}(\tau+2)\log(p\vee T)\Big) \lesssim \frac{Tp}{\sqrt{\log(p\vee T)}}\exp\Big(-(\tau+2)\log(p\vee T)\Big)=o( \frac{1}{(p\vee T)^\tau})
\end{equation}
holds for $\tau \geq 1$.
\end{lemma}
\textit{proof:}
\begin{align*}
|x_{j,k}(t)| & = |\int_{(t - \kappa_{\supp})\vee 0}^{t} \kappa_{j,k}(t - s)\d N_k(s)|\\
& \leq \kappa_{\max} \int_{t - \kappa_{\supp}}^{t}  \d N_k(s)\\
& \leq \kappa_{\max}\Big(N_k\big(T\wedge(v+2)\kappa_{\supp}\big) - N_k(v\kappa_{\supp})\Big),
\end{align*}
where $v \kappa_{\supp} \leq t - \kappa_{\supp}  < t \leq T \wedge (v+2)\kappa_{\supp}$. Thus

\begin{align*}
\mbP(\sup_{k,t\in[0,T]}|x_{j,k}(t)| > \eta) & \leq \mbP\Big(\max_{k,v}[N_k(T\wedge(v+2)\kappa_{\supp}) - N_k(v\kappa_{\supp})] > \frac{\eta}{\kappa_{\max}}\Big) \\
& \lesssim T p\frac{(h_{\max}\kappa_{\supp})^{\frac{\eta}{k_{\max}}}}{(\frac{\eta}{\kappa_{\max}})!}
e^{-h_{\max}\kappa_{\max}}\\
& \lesssim\frac{Tp}{\sqrt{\eta}}\exp\Big(\frac{\eta}{\kappa_{\max}}(1 + \log(h_{\max}\kappa_{\supp}) - \log\frac{\eta}{\kappa_{\max}})\Big)
\end{align*}
The inequality we desire holds if we take $\eta = (\tau+2)\log (p\vee T)$  for $T,p$ both large enough:
\begin{equation*}
\mbP\Big(\sup_{k,t \in [0,T]}|x_{j,k}(t)| > \kappa_{\max}(\tau+2)\log(p\vee T)\Big) \lesssim \frac{Tp}{\sqrt{\log(p\vee T)}}\exp\Big(-(\tau+2)\log(p\vee T)\Big)=o( \frac{1}{(p\vee T)^\tau})
\end{equation*}

A similar proof can show that
\begin{align*}
& \mbP\Big(\sup_{k,t \in [0,T] \ \text{and}\ \d N_{j,k}(t) = 0}|x_{j,k}'(t)| > \kappa_{\max}(\tau+2)\log(p\vee M)\Big) \\
\lesssim & \frac{Tp}{\sqrt{\log(p\vee T)}}\exp\Big(-(\tau+2)\log(p\vee T)\Big)=o( \frac{1}{(p\vee T)^\tau})
\end{align*}\hfill$\Box$

For simplicity of notation, we define the events $$C_\tau = \Big\{\sup_{k,t\in[0,T]}|x_{j,k}(t)| \leq \kappa_{\max}(\tau+2)\log(p\vee T),\ \sup_{k,t \in [0,T] \ \text{and}\ \d N_{j,k}(t) = 0}|x_{j,k}'(t)| > \kappa_{\max}(\tau+2)\log(p\vee T)\Big\}$$
and $\mbP(C_{\tau})=1 - o( \frac{1}{(p\vee T)^\tau}).$
\begin{theorem}
\label{ChenOriginal}
Suppose that $N$ is a temporal point process with intensity which satisfies those assumptions above.\ Suppose also that the functions $f_{i,k}(\cdot)$ have uniformly bounded support and
$$\sup_{i,k \in [p]}\|f_{i,k}\|_{\infty} \equiv \sup_{i,k \in [p]}\max_x|f(x)| \leq C_f.$$
\ Then,\ for $r > 0$, we have
\begin{align*}
&\mbP\bigg(\bigcup_{1 \leq i \leq k \leq p} \Big\{|\bar{y}_{i,k} - \mbE\bar{y}_{i,k}| \geq \epsilon\Big\}\bigg)\\
\leq & c_1p^2T\exp\Big(-\frac{(\epsilon T)^{r/(3r+1)}}{c_2}\Big) + p^2\exp\Big(-c_3\epsilon^2T\Big) + p^2\exp\Big[-c_4T\exp\Big(\frac{(\epsilon T)^{r(2r+1)/(3r+1)^2}}{c_5[\log(\epsilon T)]^{r/(3r+1)}}\Big)\Big].
\end{align*}
where
$$\bar{y}_{i,k} \equiv \frac{1}{T} \int_0^T \int_0^T f_{i,k}(t - t')\d N_i(t)\d N_k(t'),\ \ 1 \leq i,k \leq p,$$
$c_1,c_2,c_3,c_4,c_5$ are positive constants that do not depend on $p$ or $T$.\ Take $\epsilon = \sqrt{\frac{(\tau+2)\log(p\vee T)}{c_3T}}, r=1$ and we have
$$\mbP\bigg(\bigcup_{1 \leq i \leq k \leq p} \Big\{|\bar{y}_{i,k} - \mbE\bar{y}_{i,k}| \geq \sqrt{\frac{(\tau+2)\log(p\vee T)}{T}}\Big\}\bigg) \leq \frac{c_6}{(p\vee T)^\tau}.$$
\end{theorem}
The inequation above stems from the proof of Theorem 3 in \citeauthor{2017chen}(\citeyear{2017chen}) and $r > 0$ is introduced when making assumption about the tail of transfer kernel (see Assumption 2 of \citeauthor{2017chen}(\citeyear{2017chen}) for more details). In our setting, the transfer kernels are compact supported and we can choose any $r > 0$.

\begin{lemma}
\label{Wangxx^Tconcentration}
For $T$ large enough,\ we have constants $c_1,\ c_2,\ c_3$ that do not depend on $p$ and $T$ that
$$\mbP\bigg(\bigcup_{1 \leq i \leq k \leq p} \bigg[\bigg|\frac{1}{T}\int_0^T x_{j,i}(t)x_{j,k}(t) dt-\mbE(x_{j,i}(t) x_{j,k}(t))\bigg| \geq c_1\sqrt{\frac{(\tau+2)\log(p\vee T)}{T}} \bigg]\bigg) \leq \frac{c_2}{(p\vee T)^\tau}.$$
holds.\

\end{lemma}
\textit{Proof.}
\begin{align*}
\frac{1}{T}\int_0^T x_{j,i}(t)x_{j,k}(t) \d t & = \frac{1}{T}\int_0^T \big(\int_0^t \kappa_{j,i}(t-s)\d N_i(s)\big)\big(\int_0^T \kappa_{j,k}(t-r)dN_k(r)\big) \d t \\
& = \frac{1}{T} \int_0^T\int_0^T \Big(\int_{s\vee r}^T \kappa_{j,i}(t-s)\kappa_{j,k}(t-r) \d t\Big) \d N_i(s)\d N_k(r)
\end{align*}
Generally,\ $\int_{s\vee r}^T \kappa_{j,i}(t-s)\kappa_{j,k}(t-r) \d t$ is not a function only depend on $r - s$.\ However,\ for all $i,j \in [p]$,\ $\kappa_{j,i}(t)$ is supported in $[0,{\kappa_{\supp}}]$ and we observe that for $s\vee r < T - \kappa_{\supp}$,\ we have
$$\int_{s\vee r}^T \kappa_{j,i}(t-s)\kappa_{j,k}(t-r) \d t =
\begin{cases}
\int_0^{\kappa_{\supp}} \kappa_{j,i}(t) \kappa_{j,k}(t+s-r) \d t \ \ \ & s\geq r \\
\int_0^{\kappa_{\supp}} \kappa_{j,i}(t+r-s)\kappa_{j,k}(t)  \d t \ \ \ & s < r
\end{cases}
$$
both only depend on $s - r$.\ Thus
\begin{align*}
&\frac{1}{T}\int_0^T \int_0^T \Big(\int_{s\vee r}^T \kappa_{j,i}(t-s) \kappa_{j,k}(t-r) \d t\Big) \d N_i(s) \d N_k(t)\\
 = & \frac{1}{T}\int_0^{T-{\kappa_{\supp}}} \int_0^{T-{\kappa_{\supp}}} \Big(\int_{s\vee r}^T \kappa_{j,i}(t-s) \kappa_{j,k}(t-r) \d t\Big) \d N_i(s) \d N_k(t) \\
 + & \frac{1}{T}\int_{T-{\kappa_{\supp}}}^T \int_0^{T}\Big(\int_{s\vee r}^T \kappa_{j,i}(t-s) \kappa_{j,k}(t-r) \d t\Big) \d N_i(s) \d N_k(t) \\
 + & \frac{1}{T}\int_0^{T-{\kappa_{\supp}}} \int_{T-{\kappa_{\supp}}}^{T}\Big(\int_{s\vee r}^T \kappa_{j,i}(t-s) \kappa_{j,k}(t-r) \d t\Big) \d N_i(s) \d N_k(t) \\
 = & I^{(1)}_{ik} + I^{(2)}_{ik} + I^{(3)}_{ik}.
\end{align*}
Let
$$f_{i,k}(s-r) = \begin{cases}
\int_0^{\kappa_{\supp}} \kappa_{j,i}(t) \kappa_{j,k}(t+s-r) \d t \ \ \ & s- r \geq 0 \\
\int_0^{\kappa_{\supp}} \kappa_{j,i}(t+r-s)\kappa_{j,k}(t)  \d t \ \ \ & s- r < 0.
\end{cases}$$
With Theorem \ref{ChenOriginal},\ we have
$$\mbP\Big(\bigcup_{1 \leq i \leq k \leq p} |I^{(1)}_{ik} - \mbE[I^{(1)}_{ik}]| \geq c_3\sqrt{\frac{\tau\log(p\vee T)}{T}} \Big) \leq \frac{c_4}{(p\vee T)^\tau}.$$
As for $I^{(2)}_{ik}$ and $I^{(3)}_{ik}$,\ we have
\begin{align*}
I^{(2)}_{ik} = & \bigg|\frac{1}{T}\int_{T-{\kappa_{\supp}}}^T \int_0^{T}\Big(\int_{s\vee r}^T \kappa_{j,i}(t-s) \kappa_{j,k}(t-r) \d t\Big) \d N_i(s) \d N_k(t)\bigg| \\
\leq & \bigg|\frac{1}{T}\int_{T-{\kappa_{\supp}}}^T \int_{T-2{\kappa_{\supp}}}^{T} \Big(\int_{s\vee r}^T \kappa^2_{\max} \d t\Big)\d N_i(s) \d N_k(t)\bigg| \\
\leq & \frac{8{\kappa_{\supp}}^3\kappa^2_{\max}}{T}
\end{align*}
and
\begin{align*}
I^{(3)}_{ik} = & \bigg|\frac{1}{T}\int_0^{T-{\kappa_{\supp}}} \int_{T-{\kappa_{\supp}}}^{T}\Big(\int_{s\vee r}^T \kappa_{j,i}(t-s) \kappa_{j,k}(t-r) \d t\Big) \d N_i(s) \d N_k(t)\bigg| \\
\leq & \bigg|\frac{1}{T}\int_{T-2{\kappa_{\supp}}}^{T-{\kappa_{\supp}}} \int_{T-{\kappa_{\supp}}}^{T}\Big(\int_{s\vee r}^T \kappa^2_{\max} \d t\Big) \d N_i(s) \d N_k(t)\bigg| \\
\leq & \frac{4{\kappa_{\supp}}^3\kappa^2_{\max}}{T}.\
\end{align*}
Similarly we have
$$\mbE[I^{(2)}_{ik}] \leq \frac{8{\kappa_{\supp}}^3\kappa^2_{\max}}{T},\ \mbE[I^{(3)}_{ik}] \leq \frac{4{\kappa_{\supp}}^3\kappa^2_{\max}}{T}.$$
Thus $|I^{(2)}_{ik} - \mbE[I^{(2)}_{ik}]| + |I^{(2)}_{ik} - \mbE[I^{(2)}_{ik}]| = O(T^{-1})$.\ We have
$$\mbP\bigg(\bigcup_{1 \leq i \leq k \leq p} \bigg[\bigg|\frac{1}{T}\int_0^T x_{j,i}(t)x_{j,k}(t) dt-\mbE(x_{j,i}(t) x_{j,k}(t))\bigg| \geq c_3\sqrt{\frac{\tau\log(p\vee T)}{T}} \bigg]\bigg) \leq \frac{c_6}{(p\vee T)^\tau}.$$
\hfill$\Box$

\begin{lemma}
\label{lemmaOneOrderConcentrate}
Under Assumptions \ref{Assumption1} and \ref{Assumption2}, we have
$$\mbP\Big(\max_{k}\Big|\frac{1}{T}\int_0^Tx_{j,k}(t)\d t - \mbE[x_{j,k}(t)]\Big| = c_1\sqrt{\frac{\tau\log(p\vee T)}{T}}\Big) \leq 1 - \frac{c_2}{(p\vee T)^\tau},$$
where $c_1,c_2$ are positive constants that do not depend on $p$ or $T$.
\end{lemma}
Recall the well known Bernstein type inequalities for point process which can be found in \cite{shorack2009empirical}:
$$\mbP\Big(|M_{T}| \geq \sqrt{2vx} + \frac{Bx}{3}\ \text{and}\ \int_{0}^T g^2(t)\d \Lambda(t)\leq v \ \text{and}\ \sup_{t\in[0,T]}|g(t)| \leq B\Big) \leq 2e^{-x},$$
where $M_T = \int_0^T g(t)(\d N(t) - \d \Lambda(t))$.

Notice that
\begin{align*}
\frac{1}{T}\int_0^T (x_{j,k}(t) - \mbE[x_{j,k}(t)])\d t & = \frac{1}{T}\int_0^T\int_0^t \kappa_{j,k}(t-s)\Big(\d N_k(s) - \d \Lambda_k(s)\Big) \d t\\
& = \frac{1}{T}\int_0^T \int_s^T \kappa_{j,k}(t-s) \d t (\d N_k(s) - \d \Lambda_k(s))
\end{align*}
where the last equality is by Fubini's theorem. Let $f_1(s) = \int_s^T \kappa_{j,k}(t-s) \d t$. Since the transition kernel is bounded and supported on a compact set, $f_1(s)$ is bounded by a constant $B$ that does not depend on $p$ or $T$. Thus we have

$$\mbP\Big(\Big|\int f_1(t) (\d N_j(t) - \d \Lambda_j(t))\Big| \geq \sqrt{2B^2Tx} + \frac{Bx}{3}\ \text{and}\ \int_{0}^T f_1^2(t)\d \Lambda_j(t)\leq B^2T \ \text{and}\ \sup_{t\in[0,T]}|f_1(t)| \leq B\Big) \leq 2e^{-x}.$$

Take $x = (\tau+2)\log (p \vee T)$ and we have

$$\mbP\Big(\max_{k}\Big|\frac{1}{T}\int_0^Tx_{j,k}(t)\d t - \mbE[x_{j,k}(t)]\Big| = c_1\sqrt{\frac{\tau\log(p\vee T)}{T}}\Big) \leq 1 - \frac{c_2}{(p\vee T)^\tau},$$
\hfill$\Box$

\begin{lemma}
\label{ConcentrationOfQuadratic}
For a fixed positive constant $r > 0$ and $R > 0$, let
$$Z(r) = \sup_{(\delta_{\mu},\bDelta^\tr) \in \mbS_2(1)\cap\mbS_1(r)}\Big|\frac{1}{T}\int_0^T(\delta_\mu+\langle\bDelta,\bx_{j}(t)\rangle)^2 - \mbE(\delta_\mu+\langle\bDelta,\bx_{j}(t)\rangle)^2\Big|.$$
We have for any arbitrary constant $\tau \geq 1$,
$$\mbP\Big(Z(r) \geq c_3r^2(\tau+2)^2 \sqrt{\frac{\log(p\vee T)}{T}} \Big) \leq \frac{c_1}{(p\vee T)^\tau}.$$
\end{lemma}
\textit{proof:}
By Lemma \ref{Wangxx^Tconcentration} and \ref{lemmaOneOrderConcentrate} above, we have
$$\mbP\Big(\max_{i,k}\Big|\frac{1}{T}\int_0^T x_{j,i}(t)x_{j,k}(t) - \mbE[x_{j,i}(t)x_{j,k}(t)]\Big| > c_1(\tau+2)^2 \sqrt{\frac{\log(p\vee T)}{T}}\Big) \leq \frac{c_1}{(p\vee T)^\tau},$$

$$\mbP\Big(\max_{k}\Big|\frac{1}{T}\int_0^T x_{j,k}(t) - \mbE[x_{j,k}(t)]\Big| > c_1(\tau+2)^2\sqrt{\frac{\log(p\vee T)}{T}}\Big) \leq \frac{c_2}{(p\vee T)^\tau},$$
where $c_1$ and $c_2$ are positive constants that do not depend on $p$ or $T$.

Note that for any $(\delta_{\mu},\bDelta^\tr) \in \mbS_2(1)\cap\mbS_1(r)$, \begin{align*}
& \Big|\frac{1}{T}\int_0^T(\delta_\mu+\langle\bDelta,\bx_{j}(t)\rangle)^2 - \mbE(\delta_\mu+\langle\bDelta,\bx_{j}(t)\rangle)^2\Big|\\
 \leq & \|(\delta_{\mu},\bDelta^\tr)\|_1^2 \cdot\Big(\max_{i,k}\Big|\frac{1}{T}\int_0^T x_{j,i}(t)x_{j,k}(t) \d t - \mbE[x_{j,i}(t)x_{j,k}(t)]\Big| \vee \max_{k}\Big|\frac{1}{T}\int_0^T x_{j,k}(t) \d t - \mbE[x_{j,k}(t)]\Big|\Big). \end{align*}

Then
$$\Big\{\max_{i,k}\Big|\frac{1}{T}\int_{0}^{T} x_{j,i}(t)x_{j,k}(t) - \mbE[x_{j,i}(t)x_{j,k}(t)]\Big| \vee \max_{k}\Big|\frac{1}{T}\int_{0}^{T} x_{j,k}(t) - \mbE[x_{j,k}(t)]\Big| \leq \epsilon\Big\} \subset \{Z(r) \leq r^2\epsilon\}.$$

Take $\epsilon = c_3(\tau+2)^2 \sqrt{\frac{\log(p\vee T)}{T}}$ and we have what we want. \hfill$\Box$

\end{document}